\documentclass[aps,pra,twocolumn,showpacs,superscriptaddress]{revtex4}
\usepackage{amsfonts}

\usepackage{graphicx}
\usepackage{subfigure}
\usepackage{amsmath}
\usepackage{bm}

\def\bra#1{\mathinner{\langle{#1}|}}
\def\ket#1{\mathinner{|{#1}\rangle}}
\def\braket#1{\mathinner{\langle{#1}\rangle}}
\def\Bra#1{\left<#1\right|}
\def\Ket#1{\left|#1\right>}
{\catcode`\|=\active 
  \gdef\Braket#1{\left<\mathcode`\|"8000\let|\BraVert {#1}\right>}}
\def\BraVert{\egroup\,\mid@vertical\,\bgroup}
\usepackage{amsfonts}

\usepackage{graphicx}
\usepackage{subfigure}
\usepackage{amsmath}
\usepackage{bm}

\begin{document}

% braket.sty          Macros for Dirac bra-ket <|> notation and sets {|}
% Donald Arseneau     asnd@triumf.ca     Last modified 05-Dec-1999.
% This is free, unencumbered, unsupported software.
%
% Commands defined are:
% \bra{ }   \ket{ }   \braket{ }   \set{ }    (small versions)
% \Bra{ }   \Ket{ }   \Braket{ }   \Set{ }    (expanding versions)
% 
% The "small versions" use fixed-size brackets independent of their
% contents, whereas the "expanding versions" make the brackets and 
% vertical lines expand to envelop their contents (internally using 
% the \left and \right commands).  You should use the vertical bar
% character "|" to input any extra vertical lines.  In \Braket these
% vertical lines will expand to match the arguments, and in \Set the
% first vertical will expand this way.  E.g.,
%   \Braket{ \phi | \frac{\partial^2}{\partial t^2} | \psi }
%   \Set{ x\in\mathbf{R} | 0<{|x|}<5 }
%
% NOT defined is "\ketbra" (for projection operators) because I prefer
% \ket{ } \bra{ }.
%
% Because each definition is so small, it makes no sense to have a 
% complicated generic version for many bracket styles.  Instead, 
% you can just copy the definitions and change \langle or \rangle,
% < and > to what you like.
%
\def\bra#1{\mathinner{\langle{#1}|}}
\def\ket#1{\mathinner{|{#1}\rangle}}
\def\braket#1{\mathinner{\langle{#1}\rangle}}
\def\Bra#1{\left<#1\right|}
\def\Ket#1{\left|#1\right>}
{\catcode`\|=\active 
  \gdef\Braket#1{\left<\mathcode`\|"8000\let|\BraVert {#1}\right>}}
\def\BraVert{\egroup\,\mid@vertical\,\bgroup}
% The \mid@vertical is \vrule with ordinary TeX but \middle| in eTeX.
% We always avoid a \mathchoice in making the inner vertical lines.  
% Note that \right>, prints the same as \right\rangle but is faster.  
%
% \def\ketbra#1#2{\ket{#1}\bra{#2}}
% \def\Ketbra#1#2{\left|{#1}\vphantom{#2}\right>\left<{#2}\vphantom{#1}\right|}

% \Set{...|...} Only the first | is treated specially.
%{\catcode`\|=\active
%  \gdef\set#1{\mathinner{\lbrace\,{\mathcode`\|"8000\let|\midvert #1}\,\rbrace}}
%  \gdef\Set#1{\left\{\:{\mathcode`\|"8000\let|\SetVert #1}\:\right\}}}
%\def\midvert{\egroup\mid\bgroup}
%\def\SetVert{\egroup\;\mid@vertical\;\bgroup}

% If the user is using e-TeX with its \middle primitive, use that for
% verticals instead of \vrule.
%
%\begingroup
% \edef\@tempa{\meaning\middle}
% \edef\@tempb{\string\middle}
%\expandafter \endgroup \ifx\@tempa\@tempb
% \def\mid@vertical{\middle|}
%\else
% \let\mid@vertical\vrule
%\fi

%TCIDATA{OutputFilter=LATEX.DLL}
%TCIDATA{LastRevised=Sun Sep 24 19:45:04 2006}
%TCIDATA{<META NAME="GraphicsSave" CONTENT="32">}
%TCIDATA{CSTFile=revtx4tci.cst}

%%\input{tcilatex}

%\begin{document}

\title{Quantum Coherent Dynamics at Ambient Temperature in Photosynthetic
  Molecules}  
\author{Zachary~B.~Walters}
\affiliation{Max Planck Institute for Physics of Complex Systems, Dresden, Germany}
\email{zwalters@pks.mpg.de}

%\date{\today}

%\maketitle

%\begin{affiliations}
%\item Max Planck Institute for Physics of Complex Systems, Dresden, Germany
%\end{affiliations}

\begin{abstract}
Photosynthetic antenna complexes are responsible for absorbing energy
from sunlight and transmitting it to remote locations where it can be
stored. Recent experiments have found that this process involves
long-lived quantum coherence between pigment molecules, called
chromophores, which make up these complexes
Expected to
decay within 100 fs at room temperature, these
coherences were 
instead found to persist for picosecond time scales, despite having no
apparent isolation from the thermal environment of the cell.  This 
paper derives a quantum master equation which describes the coherent
evolution of a system in strong contact with a thermal environment.
Conditions necessary for long coherence lifetimes are identified, and
the role of coherence in efficient energy transport is illuminated.
Static spectra and exciton transfer rates for the PE545 complex of the
cryptophyte algae CS24 are calculated and shown to have good agreement with
experiment.  
\end{abstract}


\maketitle

When a solar photon is captured by a photosynthetic organism, it creates a
localized excited state, or exciton, in one of several optically active
pigment molecules, or chromophores, in a specialized antenna complex.  The
exciton is then conveyed via electrostatic coupling to a remote reaction
center where its energy can be stored and ultimately used to do work.  Because
the exciton has a natural lifetime on the order of nanoseconds, it is
necessary that the transfer process be very rapid, so that the exciton's
energy can be stored before it decays.

\begin{figure}
\includegraphics[width=\columnwidth]{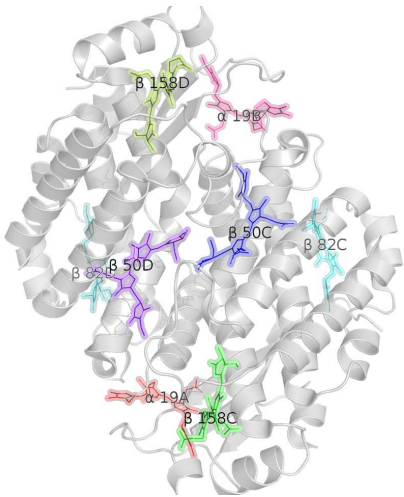}
\caption{Spatial arrangement of chromophores and protein backbone in PE545.}
\label{fig:PE545_structure}
\end{figure}

In order to understand the dynamics of photosynthesis, it is necessary to
understand the dynamics of the exciton transfer process.  However, such an
understanding is complicated by the many internal degrees of freedom in the
chromophores, and by interactions between the chromophore and the cellular
environment.  In the limit of weak coupling between chromophores, F\"orster
theory \cite{beljonne2009beyond} describes the transfer of excitons as a fully
incoherent process.  In the opposite limit, of weak coupling between
chromophores and the surrounding environment, Redfield theory
\cite{ishizaki2009adequacy} describes the coherent transport of excitons using
a perturbative expansion over a weak system-reservoir coupling, but may give
unphysical negative or diverging exciton populations.  Such unphysical
behavior may be avoided through use of a secular approximation
\cite{may2004charge}, at the cost of decoupling the evolution of populations
and coherences.  Neither the F\"orster nor the Redfield limits is applicable
to the limit in which the coupling between chromophores and between
chromophores and the environment are both strong.

Nonperturbative approaches to the problem of exciton dynamics are less easily
classified.  The Haken, Reineker, Strobl
model\cite{haken1973exactly,capek1985generalized} calculates the evolution of
an electronic reduced density matrix due to electrostatic couplings between
chromophores, with dephasing due to interaction with the environment.  This
model predicts long lived coherences between chromophores, but yields an
equilibrium state with equal occupation probabilities for all chromophores,
which is true only in the high temperature limit.  Other approaches include
the stochastic Schr\"odinger equation \cite{diósi1997non,strunz1996linear} and
methods based on quantum walks\cite{mohseni2008environment}.  Several
studies have found that interplay between the system and the environment,
including nonmarkovian effects and details of the system's vibrational
structure, can affect coherence lifetimes and the rates of exciton transfer
\cite{rebentrost2009environment,caruso2009}.

The need for an improved treatment of exciton dynamics has been highlighted by
the recent observation of long lived coherences between chromophores in
photosynthetic antenna complexes
\cite{lee2007coherence,collini2010coherently,gregory2007evidence,
  savikhin1997oscillating}.  Whereas a simple energy-time uncertainty approach
would suggest that such coherences must die in less than 100 fs
\cite{van2010quantum}, they were instead found to persist for picosecond
timescales -- on the order of the time needed for excitons to travel between
chromophores.  Rather than being confined to the low temperature limit, such
lifetimes were observed even at room temperature, shedding doubt on the common
assumption that biological systems are too "hot and wet" to display quantum
coherent behavior \cite{sarovar2010quantum}.

The observation of long lived coherences in biological systems at room
temperature raises several questions.  First, how can these coherences persist
in the thermal environment of the cell?  Are the coherences somehow protected
from interaction with the outside world, or are they preserved through another
mechanism?  Second, can the organism exploit coherence to improve the
efficiency of photosynthesis?  Does coherent transport alter the rate at which
excitons travel between chromophores?  Does the quantum information contained
in the coherence terms affect the flow of population through the complex?  
Some have speculated that the complex may perform quantum information
processing to optimize the transfer path \cite{gregory2007evidence}, although
others \cite{hoyer2010limits,rebentrost2009environment} have found that such
speedup may be short lived.
In addition to
these questions, the underlying problem of a few state system with internal
degrees of freedom interacting with a thermal reservoir may have applications
in fields such as quantum computing, where loss of coherence due to
interactions with the surrounding environment is a major limiting factor
\cite{unruh1995maintaining}.

This paper presents a new theory of coherent evolution in a multichromophore
system which does not rely on a perturbative parameter.  Instead, an
assumption of rapid thermalization of vibrational degrees of freedom is used
to derive equations of motion for a slowly evolving electronic reduced density
matrix.  In this way, both the transfer of exciton population and the
evolution of coherences are treated on an equal footing.  The resulting theory
gives a simple explanation for long coherence lifetimes in the high
temperature limit for a system which interacts strongly with its environment.

Section \ref{section:equationsofmotion} derives equations of motion for an
electronic reduced density matrix.  Part \ref{subsection:densitymatrixreduction}
reduces the full electronic + vibrational+environment density matrix to yield
equations of motion for a reduced electronic density matrix, while part
\ref{subsection:symmetrization} reduces these equations of motion to the form of the
Haken Reineker Strobl model \cite{haken1973exactly}, in which both the density
matrix and the effective coupling incorporate thermodynamic quantities.
Section \ref{section:twochromophores} applies the resulting theory to a two
chromophore system, to find lifetimes for the decay of coherence or the
transfer of population between two chromophores.  Section
\ref{section:theoryvsexpt} applies this theory to the PE545 antenna complex of
cryptophyte algae Rhodomonas CS24, shown in Figure \ref{fig:PE545_structure},
comparing to experimentally measured transfer times, as well as absorption,
circular dichroism and fluorescence spectra.

\section{Equations of motion }
\label{section:equationsofmotion}

Although the primary function of an antenna complex is to transfer electronic
excitation between chromophores, it is not a purely electronic system in the
sense of having electronic degrees of freedom which can be separated from the
other degrees of freedom in the system.  Rather, both the ground and excited
"states" of a particular chromophore correspond to an infinite number of
vibronic states.  Each of these vibronic states may interact with the local
environment of the cell, a system which is complicated and difficult to
characterize.  In order to calculate equations of motion for the purely
electronic quantity of interest, it is necessary to remove the vibrational and
environmental information from the system -- ie, to reduce the full electronic
+ vibrational + environmental density matrix to a purely electronic reduced
density matrix.

\subsection{The Vibrational/Electronic System}

The full electronic + vibrational + environmental Hamiltonian may be
partitioned into an electronic + vibrational system interacting with a
reservoir, $H=H_{S}+H_{R}+H_{SR}$.  Here the part of the total Hamiltonian
dealing with the system is relatively well characterized, while that dealing
with the reservoir is less so.  To aid in the reduction to a purely electronic
density matrix, it is useful to write vibronic states as the product of
vibrational and electronic eigenstates.  Writing the system Hamiltonian
$H_{S}=H^{\text{el}}_{0}+H^{\text{vib}}_{0}+V$, system eigenstates in the
single exciton manifold can be written $\ket{i,\vec{n}}$, where $i$ is the
excited chromophore and $\vec{n}$ the vector of excited vibrational states,
$\vec{n}=n_{1},n_{2},n_{3}\dots$.  In this basis,
$H^{\text{el}}_{0}\ket{i,n}=E_{i}\ket{i,n}$,
$H^{\text{vib}}_{0}\ket{i,n_{i}}=\epsilon^{(i)}_{n}\ket{i,n_{i}}$ and
$\epsilon^{(i)}_{\vec{n}}=\sum_{j}\epsilon^{(i)}_{n_{j}}$.  Vibrational states
of the excited chromophore $\ket{i,n_{i}}$ may differ from those of the
unexcited chromophore $\ket{0,n_{i}}$ due to differing potential energy
surfaces.  Here it is convenient to introduce electronic
$\mu^{+}_{i}=\ket{i}\bra{0}$ and vibrational
$\nu^{+}_{n_{i}}=\ket{i,n_{i}}\bra{i,0_{i}}+\ket{0,n_{i}}\bra{0,0_{i}}$
excitation operators, so that the state of the complex can be written as the
product of excitation operators acting on an initial ground state 
\begin{equation}
\psi=\sum_{i,\vec{n}}(\prod_{j} b^{j}_{m} \nu^{+}_{m_{j}})a_{i}
  \mu^{+}_{i}\ket{0}. 
\end{equation}
Deexcitation operators $\mu_{i}$ and $\nu_{n_{i}}$ are given by the Hermitian
conjugates of the excitation operators.  In the limit that electronically
exciting one chromophore does not affect the potential energy surfaces on
other chromophores, the electronic excitation/deexcitation operators commute
with the vibrational excitation/deexcitation operators for a different chromophore.

\subsection{The  reservoir}

In contrast to the system being studied, whose Hamiltonian is relatively well
known and studied, the reservoir may be complicated and difficult to
characterize.  In F\"orster theory, rapid dephasing of electronic and
vibrational degrees of freedom allow explicit consideration of the reservoir
to be avoided, as dynamics can be calculated from the overlap of the donor
emission and the acceptor absorption spectra \cite{scholes2003long}.  In
Redfield theory, the reservoir is often modeled as an infinite set of harmonic
oscillators \cite{breuer2002theory}.

%In the current work, an assumption of rapid thermalization allows electronic
%equations of motion to be found without explicit reference to the form of
%$H_{R}$ or $H_{SR}$.  As in F\"orster theory, the electronic equations of
%motion will be given in terms of an overlap between a donor emission and a
%acceptor absorption spectrum.  Along with the assumption of rapid
%thermalization, the interaction with the reservoir is incapable of
%transferring excitation between chromophores, so that
%$H_{SR}=\sum_{i}V^{(i)}_{SR}\ket{i}\bra{i}$.  

In the current work, each chromophore is assumed to interact with a separate
reservoir, so that $H_{SR}$ can be written as
$H_{SR}=\sum_{i}V^{(i)}_{SR}\ket{i}\bra{i}$, and the system-reservoir
interaction is incapable of transferring excitation between chromophores.  As
in F\"orster theory, an assumption of rapid thermalization allows electronic
equations of motion to be derived without explicit reference to the form of $V^{(i)}_{SR}$.

If eigenstates of the reservoir Hamiltonian corresponding to chromophore $i$ are given by
$H^{(i)}_{R}\ket{\zeta_{i}}=E_{\zeta_{i}}\ket{\zeta_{i}}$, then the mixing of eigenstates
$\ket{i,\vec{n},\zeta_{i}}$ is governed by $V$ and $H_{SR}$.  The assumption of
rapid thermalization is that the short time evolution of the system is given
by $U(t,t_{0}) \approx \prod_{i}U^{(i)}(t,t_{0})$, where
$U^{(i)}(t,t_{0})=\exp[-i\int_{t_{0}}^{t} V^{(i)}_{SR}(\tau)
d\tau]\ket{i}\bra{i}$.  If $\psi(t)=\sum_{i,\vec{n},\zeta_{i}}a_{i}(t) b^{(i)}_{\vec{n}}(t)
c_{\zeta_{i}}(t)\ket{i,\vec{n},\zeta_{i}}$, the electronically diagonal character of
$U$ means that $a_{i}(t)$ does not evolve, while the vibrational and
environmental degrees of freedom are driven to thermal equilibrium 
\begin{equation}
  \text{Tr}_{\zeta_{i}=\zeta'_{i}}b^{(i)}_{n}b^{*(i)}_{n'}c^{(i)}_{\zeta_{i}} 
  c^{*(i)}_{\zeta'_{i}}=\delta_{n,n'}P^{(i)}_{n}.
\label{eq:thermaldist}
\end{equation}
Here $P^{(i)}_{n}$ is the probability of vibrational state $n$ on chromophore
$i$ being excited, while $P^{(i)}_{\vec{n}}=\prod_{j}P^{(i)}_{n_{j}}$ is the
probability of a particular vector of vibrational states being excited.

Eq. \ref{eq:thermaldist} can now be used to reduce equations of motion for the
full electronic + vibrational+environment system to a purely electronic
subsystem.

\subsection{Density Matrix Reduction}
\label{subsection:densitymatrixreduction}
Electronic equations of motion can be found by tracing over environmental and
vibrational degrees of freedom.  To this end, let 
%\begin{widetext}
\begin{equation}
\begin{split}
\rho_{i\vec{n}\zeta
  j\vec{m}\eta}=a_{i}b_{\vec{n}}c_{\zeta_{i}}\ket{i,\vec{n},\zeta_{i}}\bra{j,\vec{m},\eta_{j}}a^{*}_{j}b^{*}_{\vec{m}}c^{*}_{\eta_{j}}
\times \\
e^{-i(E_{i}+\epsilon^{i}_{\vec{n}}+E_{\zeta_{i}}-E_{j}-\epsilon^{(j)}_{\vec{m}}-E_{\eta_{j}})t},
\end{split}
\label{eq:rho}
\end{equation}
be the full density matrix written with factorized coeffients and 
\begin{widetext}
\begin{equation}
\sigma_{i\vec{n}\zeta_{i}j\vec{m}\eta_{j}}=
c^{*}_{\zeta'_{i}}\ket{\zeta'_{i}}\bra{\zeta'_{i}}
(\prod_{n'_{j}}b^{*}_{n'_{j}}\ket{n'_{j}}\bra{n'_{j}}) 
\rho_{i\vec{n}\vec{\zeta}j\vec{m}\vec{\eta}}
(\prod_{m'_{j}}b_{m'_{j}}\ket{m'_{j}}\bra{m'_{j}})
c_{\eta'_{i}}\ket{\eta'_{j}}\bra{\eta'_{j}}
\label{eq:sigma}
\end{equation}
\end{widetext}
be an auxiliary matrix defined in terms of the squares of vibrational and
environmental coefficients, so that
$a_{i}a^{*}_{j}=\sum_{\vec{n},\vec{m},\zeta_{i},\eta_{j}}
\sigma_{i\vec{n}\zeta_{i}j\vec{m}\eta_{j}}$.  Equations of motion for the
electronic coefficients $a_{i}a^{*}_{j}$ can now be found by substituting
Eq. \ref{eq:thermaldist} into equations of motion for $\sigma$.

The rate of dephasing between different chromophores is found by substituting
Eq. \ref{eq:thermaldist} into Eq. \ref{eq:sigma}, yielding
\begin{equation}
a_{i}a^{*}_{j}(t)=a_{i}a^{*}_{j}(0)\sum_{\vec{n},\vec{m}}P^{(i)}_{\vec{n}} P^{(j)}_{\vec{m}}
e^{-i(E_{i}+\epsilon^{(i)}_{\vec{n}}-E_{j}-\epsilon^{(j)}_{\vec{m}})t},
\label{eq:dephasing_sums}
\end{equation}

Similarly, equations of motion for coherent evolution are found by
substituting Eq. \ref{eq:thermaldist} into the operator equation of motion for
$\sigma$ due to $H_{S}+H_{R}$, $\frac{\partial}{\partial
  t}\sigma=-i[H_{S}+H_{R},\sigma]$, yielding
\begin{equation}
\begin{split}
\frac{\partial}{\partial t}a_{i} a^{*}_{j}=-i[V_{i
  \vec{n}k\vec{p}}P^{(k)}_{\vec{p}} P^{(j)}_{\vec{m}} a_{k}a^{*}_{j}
e^{-i(E_{i}+\epsilon^{(i)}_{\vec{n}}-E_{k}-\epsilon^{(k)}_{\vec{p}})t}- \\
a_{i}a^{*}_{k}P^{(i)}_{\vec{n}} P^{(k)}_{\vec{p}} 
V_{k\vec{p}j\vec{m}}
e^{-i(E_{k}+\epsilon^{(k)}_{\vec{p}}-E_{j}-\epsilon^{(j)}_{\vec{m}})t}],
\label{eq:daijdt_sums}
\end{split}
\end{equation}
where $V_{k\vec{p}j\vec{m}}=\bra{0}\mu_{i}\nu_{\vec{n}} V
\nu^{+}_{\vec{m}}\mu^{+}_{j}\ket{0}$. 

The complicated forms of Eqs. \ref{eq:dephasing_sums} and
\ref{eq:daijdt_sums}, which result in part from their generality, may be
simplified by making assumptions about the vibrational spectrum and the form
of the vibrational/electronic operator $V$.  In terms of the vibrational
density of states
\begin{equation}
D^{(i)}(\epsilon)=\sum_{\vec{n}}\delta(\epsilon-\epsilon^{(i)}_{\vec{n}})
\end{equation}
and thermodynamic weighting factor
\begin{equation}
P^{(i)}(\epsilon)=P^{(i)}_{\vec{n}}|_{\epsilon=\epsilon^{(i)}_{\vec{n}}},
\end{equation}
the dephasing between two chromophores given in Eq. \ref{eq:dephasing_sums}
becomes 
\begin{equation}
\begin{split}
a_{i}a^{*}_{j}(t)=a_{i}a^{*}_{j}(0)\int d\epsilon \int d\epsilon' P^{(i)}(\epsilon) D^{(i)}(\epsilon)
P^{(j)}(\epsilon')D^{(j)}(\epsilon')\times \\ e^{-i(E_{i}+\epsilon-E_{j}-\epsilon')t}.
\end{split}
\label{eq:dephasing_int}
\end{equation}
while for $i=j$, unitarity of $U^{(i)}(t,t_{0})$ ensures that
$a_{i}a^{*}_{i}(t)=a_{i}a^{*}_{i}(0)$.

If the vibronic term $V$ is assumed to be purely electronic, so that 
$V=\sum_{ij}V_{ij}\mu^{+}_{i}\mu_{j}$, then 
\begin{equation}
V_{i \vec{n}j\vec{m}}=V_{ij} \bra{0}\mu_{i} \nu_{n_{i}}
\mu^{+}_{i}\nu^{+}_{m_{i}}
\ket{0}\bra{0}\nu_{n_{j}}\mu_{j}\nu^{+}_{m_{j}}\mu^{+}_{j}\ket{0}
\prod_{k \ne i,j} \delta_{n_{k},m_{k}}.
\label{eq:V_franckcondon} 
\end{equation}
where the matrix elements are Franck-Condon overlaps between vibrational
states of the excited and unexcited chromophores.

Further simplification can be achieved by replacing the oscillatory integrals
in Eq. \ref{eq:daijdt_sums} with their average over some period $\Delta t$.
As $\Delta t$ grows, this average goes to 1 if the energies in the exponential
sum to 0; 0 otherwise.  Including the effects of Lorentzian dephasing between
chromophores in this average yields an exponential linewidth, so that 
\begin{equation}
\frac{1}{\Delta t}\int dt
e^{-i(E_{i}+\epsilon^{(i)}_{\vec{n}}-E_{k}-\epsilon^{(k)}_{\vec{p}})t}
\frac{\beta^{2}}{\beta^{2}+t^{2}} \propto e^{-2\beta|E_{i}+\epsilon^{(i)}_{\vec{n}}-E_{k}-\epsilon^{(k)}_{\vec{p}}|}.
\end{equation}
Eq. \ref{eq:daijdt_sums} then becomes
\begin{equation}
\frac{\partial}{\partial t}(a_{i}a^{*}_{j})=-iV_{ik} T_{k \rightarrow
  i}(a_{k}a^{*}_{j})+i(a_{i}a^{*}_{k})V_{kj} T_{k \rightarrow j}
\label{eq:daijdt_Tij}
\end{equation}
where
\begin{equation}
\begin{split}
T_{i \rightarrow j}=\int d\epsilon \int d\epsilon' & D^{(i)}(\epsilon)
D^{(j)}(\epsilon') \\
& P^{(i)}(\epsilon)\mathbb{F}(\epsilon,\epsilon')\mathbb{L}(E_{i}+\epsilon -
E_{j}-\epsilon'),
\label{eq:Tij}
\end{split}
\end{equation} 
\begin{equation}
\mathbb{F}(\epsilon,\epsilon')=\bra{0}\mu_{i} \nu_{n_{i}}
\mu^{+}_{i}\nu^{+}_{m_{i}}
\ket{0}\bra{0}\nu_{n_{j}}\mu_{j}\nu^{+}_{m_{j}}\mu^{+}_{j}\ket{0}
|_{\epsilon=\epsilon^{(i)}_{n},\epsilon'=\epsilon^{(j)}_{m} }
\end{equation}
is the Franck Condon overlap and $\mathbb{L}(\Delta E)$ is the linewidth.

\subsection{The thermalized density matrix}
\label{subsection:symmetrization}
%In Eq. \ref{eq:daijdt_Tij}, $T_{i \rightarrow j}$ plays the role of the
%overlap between the donor emission and the acceptor absorption spectra in
%F\"orster theory: it represents the weighted number of acceptor states on
%chromophore $j$ which are nearly degenerate with occupied donor states on
%chromophore $i$.  
In Eq. \ref{eq:daijdt_Tij}, $T_{i \rightarrow j}$ plays the role of the
overlap between donor emission and acceptor absorption spectra in F\"orster
theory: it reflects the number of acceptor states sufficiently close in energy
to an occupied donor state to receive population.  The asymmetry of $T_{i
  \rightarrow j}$ reflects irreversible flow of population from high- to low
energy chromophores: for $E_{i}> E_{j}$, every potential donor state on
chromophore $i$ has a degenerate acceptor state on chromophore $j$, while the
reverse is true only if $\epsilon^{(j)} \ge E_{i}-E_{j}$.

The asymmetric form of Eq. \ref{eq:daijdt_Tij} which results from this
irreversible flow can be rectified by defining a symmetrized density matrix 
$\tilde{\rho}_{ij} \equiv
e^{\alpha_{i}+\alpha_{j}}a_{i}a^{*}_{j}$ and effective coupling
$\tilde{V}_{ij}=V_{ij}T_{i \rightarrow j}e^{\gamma_{i}-\gamma_{j}}$.
Requiring that the effective coupling be Hermitian yields 
$\frac{T_{i \rightarrow j}}{T_{j \rightarrow i}}=e^{2
  \gamma_{i}-2\gamma_{j}}$, while requiring that Eq. \ref{eq:daijdt_Tij} be
unaffected by the change of variables requires that $\alpha_{i}=-\gamma_{i}$.
Performing this symmetrization and approximating the dephasing between
chromophores as
exponential yields a master equation of the form
\begin{equation}
\frac{\partial}{\partial t}\tilde{\rho}=-i[\tilde{V},\tilde{\rho}]-(1-\delta_{ij})\Gamma
\tilde{\rho}. 
\label{eq:dpdt}
\end{equation}
For a constant density of states and Boltzmann probability distribution, the
Lorentzian dephasing is well approximated by setting $\Gamma=k_{B}T$.

A byproduct of this symmetrization procedure is that $\tilde{\rho}_{ij}$ now
incorporates information about the equilibrium distribution.  In
Eq. \ref{eq:daijdt_Tij}, with all coherence terms set to 0, equilibrium is
given by the condition of detailed balance, when
\begin{equation}
\frac{a_{i}a^{*}_{i}}{a_{j}a^{*}_{j}}=\frac{T_{j \rightarrow i}}{T_{i \rightarrow j}}
\end{equation}
for all $i,j$ pairs.  Due to the symmetrization procedure, this condition is
satisfied when $\tilde{\rho}$ is proportional to the identity matrix.  In this
way, the current approach differs significantly from that of Haken, Reineker
and Strobl \cite{haken1973exactly,capek1985generalized}, which applies a
quantum master equation of the form of Eq. \ref{eq:dpdt} to an unsymmetrized
density matrix, yielding equilibrium distributions in which all chromophores
have equal occupations, regardless of excitation energy or ambient
temperature.  If the density of states is again assumed to be constant and
$P^{(i)}(\epsilon)$ to be Boltzmann, $\alpha_{i}=-\gamma_{i}=\beta E_{i}/2$,
so that at equilibrium, $a_{i}a^{*}_{i}=\frac{e^{-\beta
    E_{i}}}{\sum_{j}e^{-\beta E_{j}}}$.

In the high temperature limit, where the weighting factors approach unity, the
thermalized and unthermalized density matrices approach one another.  For this
reason, Eq. \ref{eq:dpdt} may be thought of as a generalization of the HRS
model which describes dynamics at low temperatures.  Due to the identical
form, many results of the HRS model will carry through unchanged to describe
the dynamics of the thermalized density matrix.

A second consequence of the symmetrization procedure is that thermodynamic
weighting factors have been incorporated into the effective coupling.
Assuming a constant density of vibrational states with Boltzmann weighting
factors and approximating the linewidth and Franck Condon factors as
$\mathbb{F}(\epsilon,\epsilon')\mathbb{L}(E_{i}+\epsilon -
E_{j}-\epsilon')=\delta(E_{i}+\epsilon - E_{j}-\epsilon')$ yields
$\tilde{V}_{ij}=V_{ij}e^{-\beta|E_{i}-E_{j}|/2}$, so that the effective
coupling strength decays exponentially with the difference in excitation
energies at a rate dependent on the temperature.  Because $k_{B}T$ is only 200
wavenumbers at 300K, the effective interaction between two chromophores will
decrease rapidly even for small separations in excitation energy.  Because of
this, the effective coupling matrix $\tilde{V}$ will be both weaker and
effectively more sparse than the matrix of bare electronic couplings, $V$.

\section{The two chromophore system}
\label{section:twochromophores}
The equations of motion derived in the previous section give a simple
explanation for the long coherence lifetimes observed in
\cite{lee2007coherence,collini2010coherently,gregory2007evidence,
  savikhin1997oscillating}.  For a system of two chromophores, with
$\tilde{V}=\left( \begin{array}{cc}\Delta E & d \\ d & -\Delta E \end{array}
\right)$, applying Eq. \ref{eq:dpdt} twice yields differential equations for
density matrix components
\begin{align*}
&\frac{d^{2}}{dt^{2}}(\tilde{\rho}_{00}-\tilde{\rho}_{11})+\Gamma
\frac{d}{dt}(\tilde{\rho}_{00}-\tilde{\rho}_{11})+4
d^{2}(\tilde{\rho}_{00}-\tilde{\rho}_{11})=0 \\
&\frac{d^{2}}{dt^{2}}(\tilde{\rho}_{10}-\tilde{\rho}_{01})+\Gamma
\frac{d}{dt}(\tilde{\rho}_{10}-\tilde{\rho}_{01})+4
d^{2}(\tilde{\rho}_{10}-\tilde{\rho}_{01})=0 \\
&\frac{d}{dt}(\tilde{\rho}_{10}+\tilde{\rho}_{01})=-\Gamma(\tilde{\rho}_{10}
+\tilde{\rho}_{01}) \\
&\frac{d}{dt}(\tilde{\rho}_{00}+\tilde{\rho}_{11})=0.
\end{align*}

Here the total population remains constant, while the sum of off diagonal
elements decays as $e^{-\Gamma t}$, where $\Gamma=(25 \text{fs})^{-1}$ at
300K.  However, both the population imbalance and the difference of off
diagonal elements behave as damped harmonic oscillators.  If 
If $P=(\tilde{\rho}_{00}-\tilde{\rho}_{11})$ or
$(\tilde{\rho}_{10}-\tilde{\rho}_{01})$,
$P(t)=Ae^{\lambda_{+}t}+Be^{\lambda_{-}t}$, with 
\begin{equation}
\lambda_{\pm}=\frac{-\Gamma \pm \sqrt{\Gamma^{2}-16d^{2}}}{2}.
\label{eq:lambdapm}
\end{equation}
In the underdamped limit when $\Gamma < 4d$, $P(t)$ oscillates within the
exponential envelope $e^{-\Gamma t/2}$.  When $\Gamma >> 4d$, the system is
overdamped and decays without oscillating at two rates, with $\lambda_{+}
\rightarrow -4d^{2}/\Gamma$, and $\lambda_{-} \rightarrow -\Gamma$.

Significantly, long coherence times do not depend on isolation of the system
from the environment or a slow rate of dephasing $\Gamma$.  On the contrary,
long coherence times are obtained in both the underdamped limit, when $\Gamma
\rightarrow 0$ and in the overdamped limit when $\Gamma >> 4d$.  Because of
this, long coherence times will be observed in both the high and low
temperature limits.

The preceding analysis has been previously derived in the context of the HSR
model \cite{haken1973exactly,capek1985generalized}.  However, the
thermodynamic weighting of the coupling matrix elements is crucial to
accurately describing the timescale for exciton transport or the decay of
coherence.  The exponential weighting term $T_{i \rightarrow
  j}e^{\gamma_{i}-\gamma_{j}}$ which multiplies the bare matrix element
$V_{ij}$, can easily change a particular chromophore pair from the under- to
the overdamped limit.  For a constant density of states and Boltzmann
weighting factors, this term evaluates to $e^{-\beta|E_{i}-E_{j}|/2}$, so that
chromophores which are distant in energy relative to the ambient temperature
will have very weak effective couplings.

A final observation about the two chromophore system is that extremely long
coherence times may correspond to slow exciton transport through the antenna
complex.  Because the population imbalance
$(\tilde{\rho}_{00}-\tilde{\rho}_{11})$ and the antisymmetric off diagonal
component $(\tilde{\rho}_{10}-\tilde{\rho}_{01})$ obey the same differential
equation, long coherence times correspond to slow decay of population
imbalances.  As the differential equations decouple from one another, this is
not an example of coherence affecting exciton transport; rather, coherence and
population dynamics are parametrically varying with respect to the same
parameters.

\section{Comparison With Experiment -- PE545}
\label{section:theoryvsexpt}
The principles which govern the evolution of the two chromophore system apply
as well to the case of a photosynthetic complex, with the exception that, with
more than two chromophores, there are now multiple pathways for the exciton to
follow through the complex.  The effect of these pathways is limited, however,
by the thermodynamic weighting of the effective coupling, which may make the
effective coupling matrix both weaker and effectively more sparse than the
bare electronic coupling matrix.

In order to test the theory derived in Section
\ref{section:equationsofmotion}, Eq. \ref{eq:dpdt} was used to describe
dynamics in the PE545 antenna complex of the cryptophyte algae Rhodomonas
CS24, which was found to display long coherence lifetimes in
\cite{collini2010coherently}.

Because the effective coupling strength is exponentially dependent on the
energy spacing between two chromophores, it is important to use accurate
excitation energies in order to yield the correct dynamics.  Here the error in
Hartree Fock or ci singles calculations, which may be sizeable fractions of an
electron volt, may prove unacceptably large.  For this reason, the
effective couplings and site energies for PE545 were taken from
\cite{novoderezhkin2010excitation}, where the couplings were calculated ab
initio, but the site energies were found by matching to several static and
dynamic spectra, as calculated using Redfield theory.  The theory was then
tested by comparing calculated lineshapes and times for excitons to transfer
between chromophores to those measured in experiment.  Here it must be noted
that the procedure of finding excitation energies by matching to excitation
spectra may yield an artificially good agreement for lineshapes calculated
using a different theory but the same excitation energies.

\begin{figure}
\includegraphics[width=\columnwidth]{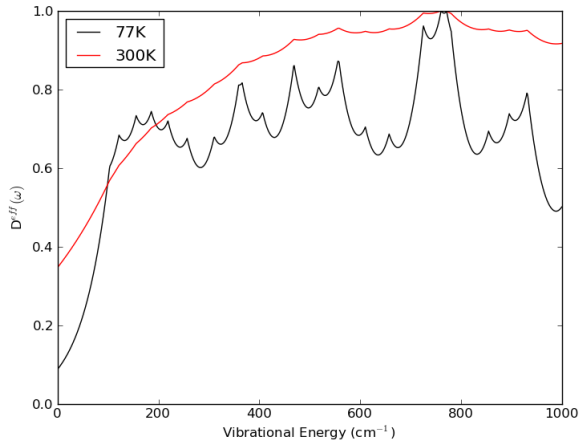}
\caption{Effective vibrational density of states, calculated for PE545
  assuming linewidth $\mathbb{L}(\Delta E)=e^{-\beta|\Delta E|}$.}
\label{fig:Deff_vs_temp}
\end{figure}

The validity of the assumptions made in deriving Eq. \ref{eq:dpdt} -- namely,
the assumption of a continuous vibrational density of states and a delta
function linewidth, can be tested by constructing an effective density of
states, corresponding to a discrete vibrational spectrum with a finite
linewidth caused by dephasing between chromophores.  Figure
\ref{fig:Deff_vs_temp} shows the effective density of states
\begin{equation}
D^{\text{eff}}(\omega)=\sum_{n,i}e^{-\beta|\omega-\omega_{i}(n+1/2)|},
\end{equation}
calculated at $t=77K$ and $300K$ for the set of vibrational energies taken
from \cite{novoderezhkin2010excitation}.  At both temperatures, the effective
density of states levels out at vibrational energies of a few hundred
wavenumbers.  At both temperatures, the effective density of states shows a
dip at zero energy, due to the zero point energy of the vibrational modes.
This drop is more accentuated at low temperatures, due to the narrowness of
the resulting linewidth.

Because of this, the assumption of a constant density of states may break down
for pairs of chromophores which are close to degenerate, or at low
temperatures.  In these limits, $T_{i \rightarrow j}$ should be calculated
using a finite summation in place of the integrals in Eq. \ref{eq:Tij}.

The description of dynamics in PE545 was tested by calculating lineshapes for
absorption, circular dichroism and fluorescence spectra, as well as the times
necessary to transfer excitation between pairs of chromophores.

PE545 consists of six phycoerythrobilin (PEB) and two dihydrobiliverdin (DBV)
chromophores, held in place by a dimer of two $\alpha\beta$ monomers.  Each
monomer contains three PEB chromophores on the $\beta$ subunit and one DBV
chromophore on the $\alpha$ subunit.  The DBV chromophores are redshifted with
respect to the PEB chromophores, with absorption maxima at 569 nm, compared to
545 for the PEB chromophores \cite{van2006energy}.  Fig
\ref{fig:PE545_structure} shows the spatial arrangement of the chromophores in
the complex.  Excitons escape the comples by first making their way to the DBV
bilins, then slowly settling onto the lowest energy DBV bilin before escaping
the complex.  Fluorescence experiments \cite{doust2006photophysics} indicate
that excitons primarily escape the complex via a single DBV bilin.  Due to
spectral overlap in the DBV and PEB bands, it may be difficult for
spectroscopic experiments to distinguish precisely which chromophores are
excited.

\subsection{Static Lineshapes -- Absorption, Fluorescence and Circular Dichroism}
Spectra for absorption, circular dichroism and fluorescence can be calculated
from time dependent expectation values of a density matrix which has been
supplemented by addition of the ground state \cite{renger2002relation}.  Here
the quantities of interest, $(a_{i}a^{*}_{j})(t)$ can be found by removing the
thermodynamic weighting factors from the thermalized density matrix, in which
the dynamics are calculated.  The absorption spectrum is given by the Fourier
transform of the dipole correlation function
\begin{equation}
d(t)=<\vec{\mu}(t)\vec{\mu}(0)>.
\end{equation}
where the dipole operator $\vec{\mu}=\sum_{i}\vec{\mu_{i}}(\ket{i}\bra{0}+\ket{0}\bra{i})$.  In
terms of the density matrix, this is given by
\begin{equation}
d(t)=\text{Tr}(\vec{\mu} \vec{\rho}(t)),
\end{equation}
where $\vec{\rho}(0)=\vec{\psi(0)}\bigotimes\vec{\psi^{*}}(0)$,
$\vec{\psi(0)}=\ket{0}+\epsilon \sum_{i}\vec{\mu_{i}}\ket{i}$ and
$\vec{\rho}(t)$ is found by propagating $\vec{\tilde{\rho}}$ according to
Eq. \ref{eq:dpdt}.  The parameter $\epsilon$ factors out of the Fourier
transform and is discarded upon normalization of the spectrum.  Dephasing
between the ground and excited states is treated as an exponential decay term
$\Gamma_{0i}=(400 cm^{-1})$, taken from \cite{novoderezhkin2010excitation}.

\begin{figure}
\includegraphics[width=\columnwidth]{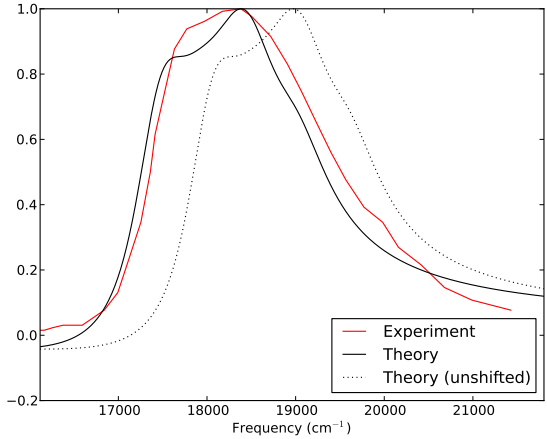}
\includegraphics[width=\columnwidth]{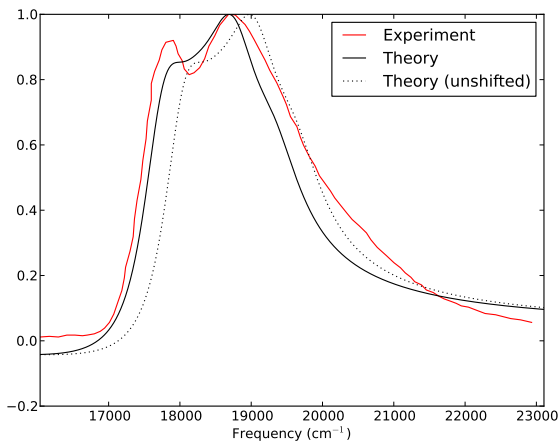}
\caption{Normalized absorption spectra $\omega*d(\omega)$ for PE545, compared
  to experimental values taken from \cite{novoderezhkin2010excitation}.  a) 300K, b)77K.}
\label{fig:PE545_absorption}
\end{figure}

\begin{figure}
\includegraphics[width=\columnwidth]{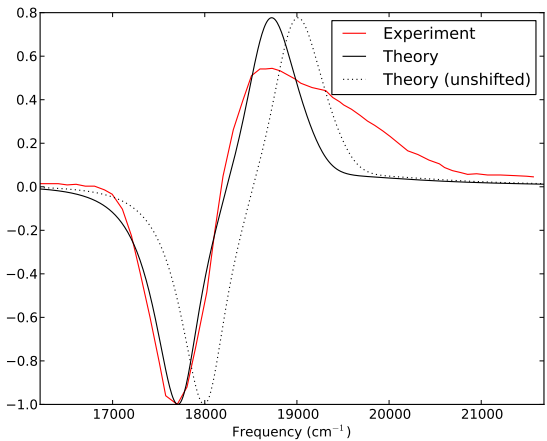}
\caption{Normalized circular dichroism spectra $\omega*m(\omega)$ 
  calculated at 300K compared to experimental values taken from
  \cite{novoderezhkin2010excitation}.} 
\label{fig:PE545_circulardichroism}
\end{figure}

In the same way, the circular dichroism spectrum is found by taking the
Fourier transform of 
\begin{equation}
\vec{m}(t)=<\vec{m}(t)\vec{\mu}(0)>
\end{equation}
where $\vec{m}=\vec{\mu}\times\vec{R}$ is the magnetic dipole operator.
Figures \ref{fig:PE545_absorption} and \ref{fig:PE545_circulardichroism}
compare the calculated absorption and circular dichroism lineshapes to those
measured in experiment in \cite{novoderezhkin2010excitation}.  Because the
current theory does not account for reorganization energy of the chromophores
following excitation, these spectra have been shifted in energy so that the
peak of the calculated spectrum aligns with the peak of the experimental
spectrum.  

\begin{figure}
\includegraphics[width=\columnwidth]{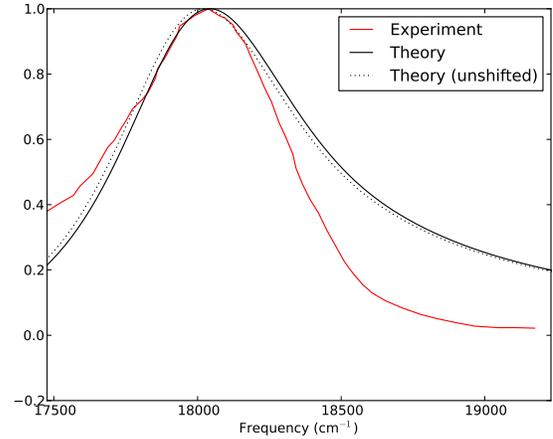}
\includegraphics[width=\columnwidth]{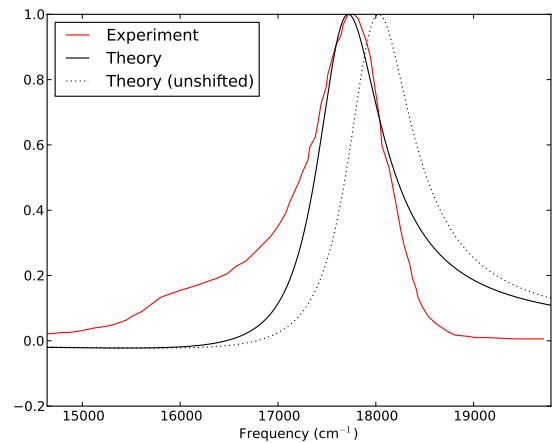}
\caption{Normalized fluorescence spectra $\omega^{3}*d(\omega)$ for PE545,
  compared to experimental values taken from \cite{novoderezhkin2010excitation}.  a) 300K, b)77K.}
\label{fig:PE545_fluorescence}
\end{figure}

In contrast to the absorption and circular dichroism spectra, the fluorescence
spectrum reflects emission at large times, for a system which has settled into
the lowest energy eigenstate.  If this eigenstate is given by $\psi_{e}$ and
$\psi(0)=\psi_{e}+\epsilon\ket{0}$, the fluorescence spectrum is proportional
to $w^{3}|\vec{f}(\omega)|^{2}$, where
\begin{equation}
\vec{f}(t)=<\vec{\mu}\rho(t)>
\end{equation}
and $\rho(0)=\psi(0)\bigotimes\psi(0)$.  Figure \ref{fig:PE545_fluorescence}
compares the normalized fluorescence spectrum to that measured in experiment.

At 300K, the calculated lineshapes show good agreement.  The width of the
absorption and the negative lobe of the circular dichroism spectra agree
closely with the experimental widths, although the positive lobe of the
circular dichroism spectrum has a broader tail and less accentuated peak than
the calculated lobe.  The calculated fluorescence spectrum agrees with
experiment in the peak region and on the blue side of the maximum, but falls
off more quickly than experiment on the red side of the maximum.

Agreement with experiment is somewhat worse at 77K.  The absorption spectrum
gives the correct width in the peak region, but falls off more quickly than
experiment on the red side of the maximum and less quickly than experiment on
the blue side.  The fluorescence spectrum shows good agreement with the
experimental lineshape on the blue side of the maximum, but falls off much
more quickly than the experiment on the red side.  Both the absorption and
fluorescence spectra miss a broad tail on the red side of the fluorescence.
The circular dichroism spectrum at 77K was not available for comparison.  

\subsection{Exciton Transfer Times}
While static lineshapes incorporate some dynamical information in the form
of a Fourier transform of a time dependent correlation function, this
information is obscured somewhat by the thermodynamic factors favoring
population of low energy chromophores and by finite linewidths due to the
decay of coherence.  In \cite{doust2005mediation},timescales for exciton
transfer were measured more directly by finding Evolution Associated
Difference Spectra (EADS) associated with particular transfer lifetimes
following an excitation pulse of a particular frequency.  Here interpretation
is complicated by spectral overlap between chromophores, and by the difficulty
in distinguishing between similar transfer lifetimes, so that only a few,
picosecond scale lifetimes could be measured, and the chromophores being
populated/depopulated could not be completely identified. 

Within these limitations, the calculated rates for exciton transfer show good
agreement with experiment.  Table \ref{table:transfertimes_measured} shows the
experimentally measured EADS transfer times, measured at 300K and 77K, for
excitation wavelengths of 485 nm and 530 nm.  Tables
\ref{table:lambdaslow_300K} and \ref{table:lambdaslow_77K} show experimental
decay times $\lambda_{+}^{-1}$, calculated using Eq. \ref{eq:lambdapm}.

\begin{figure}
\begin{tabular*}{\textwidth}{c c c c c c c}
T (k) & $\lambda_{\text{excitation}}$ (nm) & T1 (fs) & T2 (fs) & T3 (ps) & T4 (ps) &
T5 (ns) \\
298 & 485 & 25 & 250 & 1.84 & 23.5 & 1 \\
298 & 530 & 25 & 250 & 1.84 & 16.4 & $>1$ \\
77 & 485 & 25 & 960 & 3.0 & 30 & $>1$ \\
77 & 530 & 25 & 960 & 3.0 & 17.7 & $>1$ \\
\end{tabular*}
\caption{Experimentally measured transfer times, taken from
  \cite{doust2005mediation}.}
\label{table:transfertimes_measured}
\end{figure}

%\begin{widetext}
\begin{figure}
\begin{tabular*}{\textwidth}{c c c c c c c c c}
& $\beta_{50C}$ & $\alpha_{19A}$ & $\alpha_{19B}$ & $\beta_{82C}$ &
$\beta_{158C}$ & $\beta_{50D}$ & $\beta_{82D}$ & $\beta_{158D}$ \\
$\beta_{50C}$ &   &   &   &   & 3.7 &   &   &   \\ $\alpha_{19A}$ &   &   &
8.4 & 0.96 &   &   &   &   \\ $\alpha_{19B}$ &   & 8.4 &   & $0.06 + 0.09i$
&   &   &   &   \\ $\beta_{82C}$ &   & 0.96 & $0.06 + 0.09i$ &   &   &   &
&   \\ $\beta_{158C}$ & 3.7 &   &   &   &   &   &   &   \\ $\beta_{50D}$ &   &
&   &   &   &   &   &   \\ $\beta_{82D}$ &   &   &   &   &   &   &   & 7.7 \\
$\beta_{158D}$ &   &   &   &   &   &   & 7.7 &   \\

%& $\beta_{50C}$ & $\alpha_{19A}$ & $\alpha_{19B}$ & $\beta_{82C}$ &
%$\beta_{158C}$ & $\beta_{50D}$ & $\beta_{82D}$ & $\beta_{158D}$ \\
%$\beta_{50C}$ &   &   &   &   & 3.7 &   &   &   \\ $\alpha_{19A}$ &   &   &
%8.44 & 0.957 &   &   &   &   \\ $\alpha_{19B}$ &   &  &   & 0.061 + 0.091i
%&   &   &   &   \\ $\beta_{82C}$ &   &  &  &   &   &   &
%&   \\ $\beta_{158C}$ &  &   &   &   &   &   &   &   \\ $\beta_{50D}$ &   &
%&   &   &   &   &   &   \\ $\beta_{82D}$ &   &   &   &   &   &   &   & 7.69 \\
%$\beta_{158D}$ &   &   &   &   &   &   &  &   \\

\end{tabular*}
\caption{Transfer / coherence decay times (ps) shorter than 100 ps for PE545
  calculated at 77K.} 
\label{table:lambdaslow_77K}
\end{figure}
%\end{widetext}

\begin{figure}
\begin{tabular*}{\textwidth}{c c c c c c c c c}
  & $\beta_{50C}$ & $\alpha_{19A}$ & $\alpha_{19B}$ & $\beta_{82C}$ & $\beta_{158C}$ & $\beta_{50D}$ & $\beta_{82D}$ & $\beta_{158D}$ \\ $\beta_{50C}$ &   &   & 2.93 & 2.11 & 1.21 & 4.81 & 12.9 & 14.9 \\ $\alpha_{19A}$ &   &   & 20.4 & 2.64 & 7.37 &   & 19.3 &   \\ $\alpha_{19B}$ & 2.93 & 20.4 &   & 0.158 &   &   &   & 27.2 \\ $\beta_{82C}$ & 2.11 & 2.64 & 0.158 &   &   &   &   &   \\ $\beta_{158C}$ & 1.21 & 7.37 &   &   &   & 53.5 & 28.7 & 25.3 \\ $\beta_{50D}$ & 4.81 &   &   &   & 53.5 &   & 2.11 & 7.75 \\ $\beta_{82D}$ & 12.9 & 19.3 &   &   & 28.7 & 2.11 &   & 8.67 \\ $\beta_{158D}$ & 14.9 &   & 27.2 &   & 25.3 & 7.75 & 8.67 &   \\
\end{tabular*}
\caption{Transfer / coherence lifetimes (ps) shorter than 100 ps for PE545
  calculated at 300K.} 
\label{table:lambdaslow_300K}
\end{figure}

As with static spectra, agreement between theory and experiment is somewhat
better at 300K than at 77K.  At 300K, population of a DBV chromophore is found
to occur in 150 fs, vs an experimentally measured 250 fs, while transfer of
population between DBV chromophores is calculated to occur in 20.4 ps,
compared to an experimental value of 23.4 ps or 16.4 ps, depending on the
excitation frequency.  Several pairs of chromophores are calculated to have
transfer times in the range of 1-3 ps, comparable to the remaining
experimental rate of 1.84 ps.

Agreement with experiment is considerably worse at 77K.  Population of a DBV
chromophore is now calculated to occur in 956 fs, compared to an experimental
value of 960 fs, while a 3.7 ps time for transfer between $\beta_{50C}$ and
$\beta_{158C}$ matches well with a 3 ps experimental transfer times.  However,
the 8.4 ps calculated for transfer between DBV chromophores and the 7.7 ps
calculated between $\beta_{50D}$ and $\beta_{82D}$ do not correspond well with
experimental times of 30 ps and 17.7 ps.  Finally, the theoretical
calculations include an underdamped transfer between $\alpha_{19B}$ and
$\beta_{82C}$ which is not seen in the experiment, perhaps because the real
part of the transfer time, at 60 fs, is shorter than the experimental
resolution of 120 fs.

\subsection{Density Matrix Propagation vs. Pairwise Decay Model}

\begin{figure}
\includegraphics[width=\columnwidth]{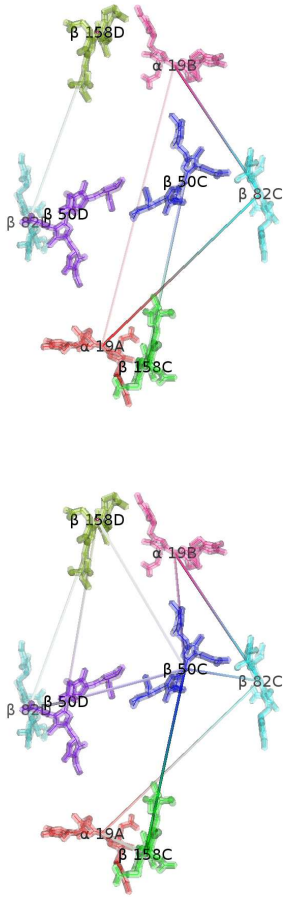}
\caption{Rates of population transfer / coherence decay in PE545.  Lines
  connecting chromophores have opacity proportional to $\lambda_{+}^{-1}$,
  with full opacity corresponding to $\lambda_{+}^{-1}=1$ ps. a) Rates
  calculated at 77K b) Rates calculated at 300K.}
\label{fig:transferrates_PE545}
\end{figure}

The agreement between the calculated pairwise decay times and transfer times
measured in experiment raise the question of which information in
$\tilde{\rho}$ is necessary to describe the dynamics of excitons in the
complex.  In principle, the entire density matrix is necessary to describe
these dynamics, as excitons may follow any pathway through the complex, so
that the amplitude to be on any particular chromophore is the sum of many
interfering pathways.  However, due to thermodynamic weighting of the
effective coupling matrix, many of these pathways are strongly overdamped,
with weak effective coupling and slow decay of any population imbalance.
Because of this, the network of efficient transfer pathways is relatively
sparse.  This can be seen in Tables \ref{table:lambdaslow_300K} and
\ref{table:lambdaslow_77K}, or pictorially in Figure
\ref{fig:transferrates_PE545}, where the opacity of the line connecting two
chromophores reflects the rate of decay for population imbalances
$(\tilde{\rho}_{ii}-\tilde{\rho}_{jj})$ (or antisymmetric coherence terms
$(\tilde{\rho}_{ij}-\tilde{\rho}_{ji})$) between them.  As a result of this
sparsity, there are relatively few efficient pathways through the complex
which may interfere with each other effectively.  In Figure
\ref{fig:transferrates_PE545}, two interfering pathways would form a closed
curve of lines with nearly equal opacity.

At 77K, $\alpha_{19A}$, $\beta_{82C}$ and $\alpha_{19B}$ form the only closed
curve, while none are present at 300K (the 150 fs transfer time between
$\alpha_{19B}$ and $\beta_{82C}$ is much faster than the ~2 ps transfer times
between $\alpha_{19B},\beta_{158C}$ and $\beta_{82C},\beta_{158C}$).

Due to the lack of interfering pathways, the flow of excitons through the
PE545 antenna complex is well approximated by a minimal model, in which
population imbalances in the thermalized density matrix decay exponentially as
\begin{equation}
\frac{d}{dt}(\tilde{\rho}_{ii}-\tilde{\rho}_{jj})=-\lambda_{+}(\tilde{\rho}_{ii}-\tilde{\rho}_{jj}),
\end{equation}
with $\lambda_{+}$ given by Eq. \ref{eq:lambdapm}.

\begin{figure}
\includegraphics[width=\columnwidth]{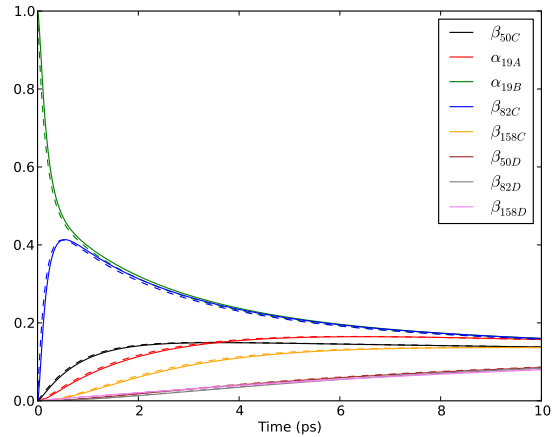}
\includegraphics[width=\columnwidth]{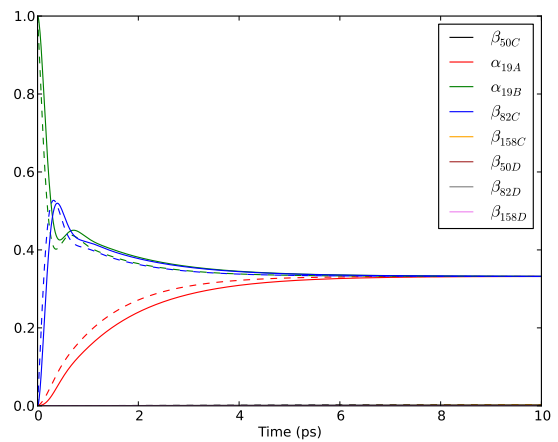}
\caption{Exciton populations vs time calculated at a) 300K b) 77K for initial
  state $\alpha_{19B}$.  Solid
  lines give populations calculated by Eq. \ref{eq:dpdt}, while dashed lines
  give populations calculated using the minimal exponential decay model.}
\label{fig:pops_vs_time_a19b}
\end{figure}

\begin{figure}
\includegraphics[width=\columnwidth]{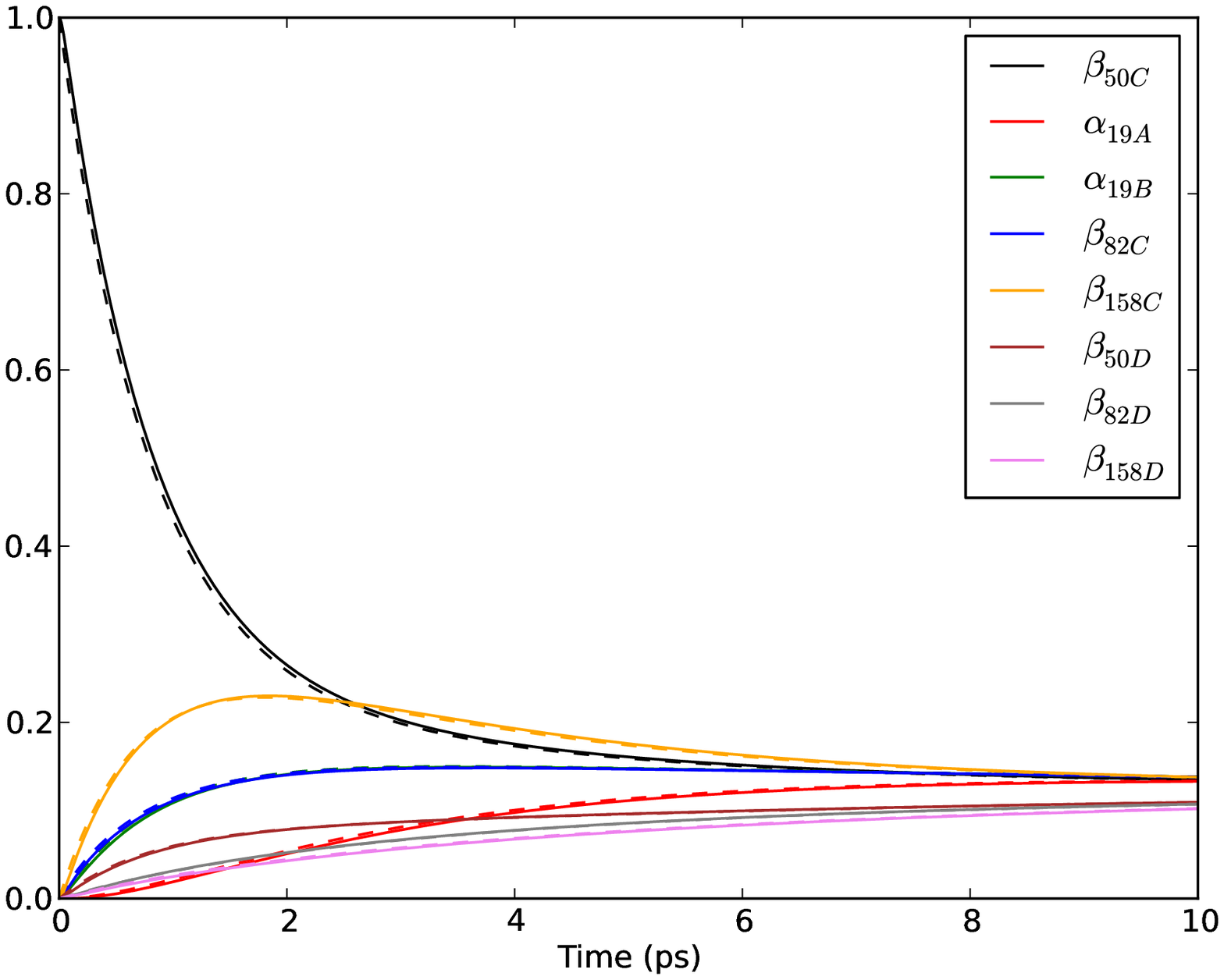}
\includegraphics[width=\columnwidth]{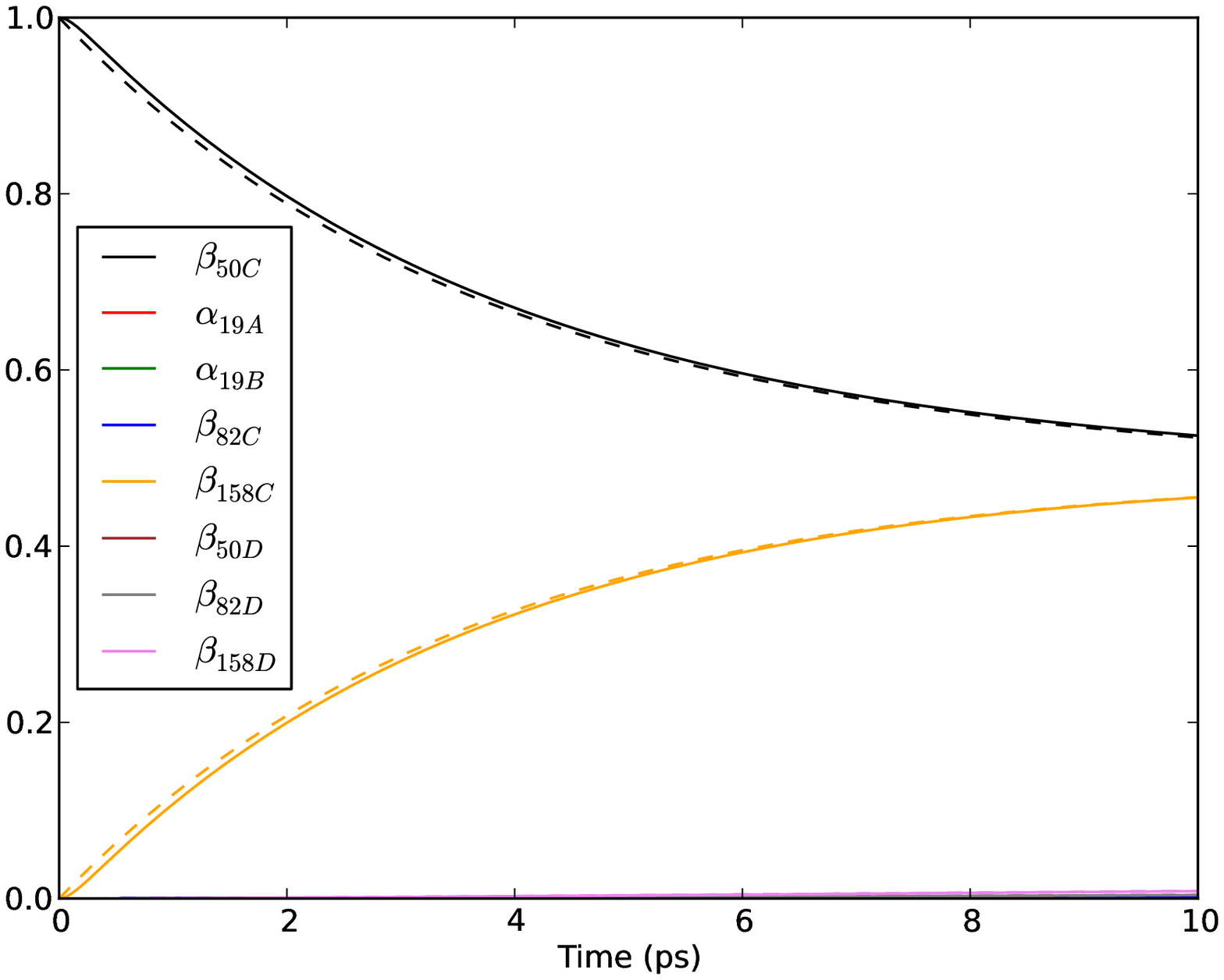}
\caption{Exciton populations vs time calculated at a) 300K b) 77K for initial
  state $\beta_{50C}$.  Solid
  lines give populations calculated by Eq. \ref{eq:dpdt}, while dashed lines
  give populations calculated using the minimal exponential decay model.}
\label{fig:pops_vs_time_b50c}
\end{figure}

\begin{figure}
\includegraphics[width=\columnwidth]{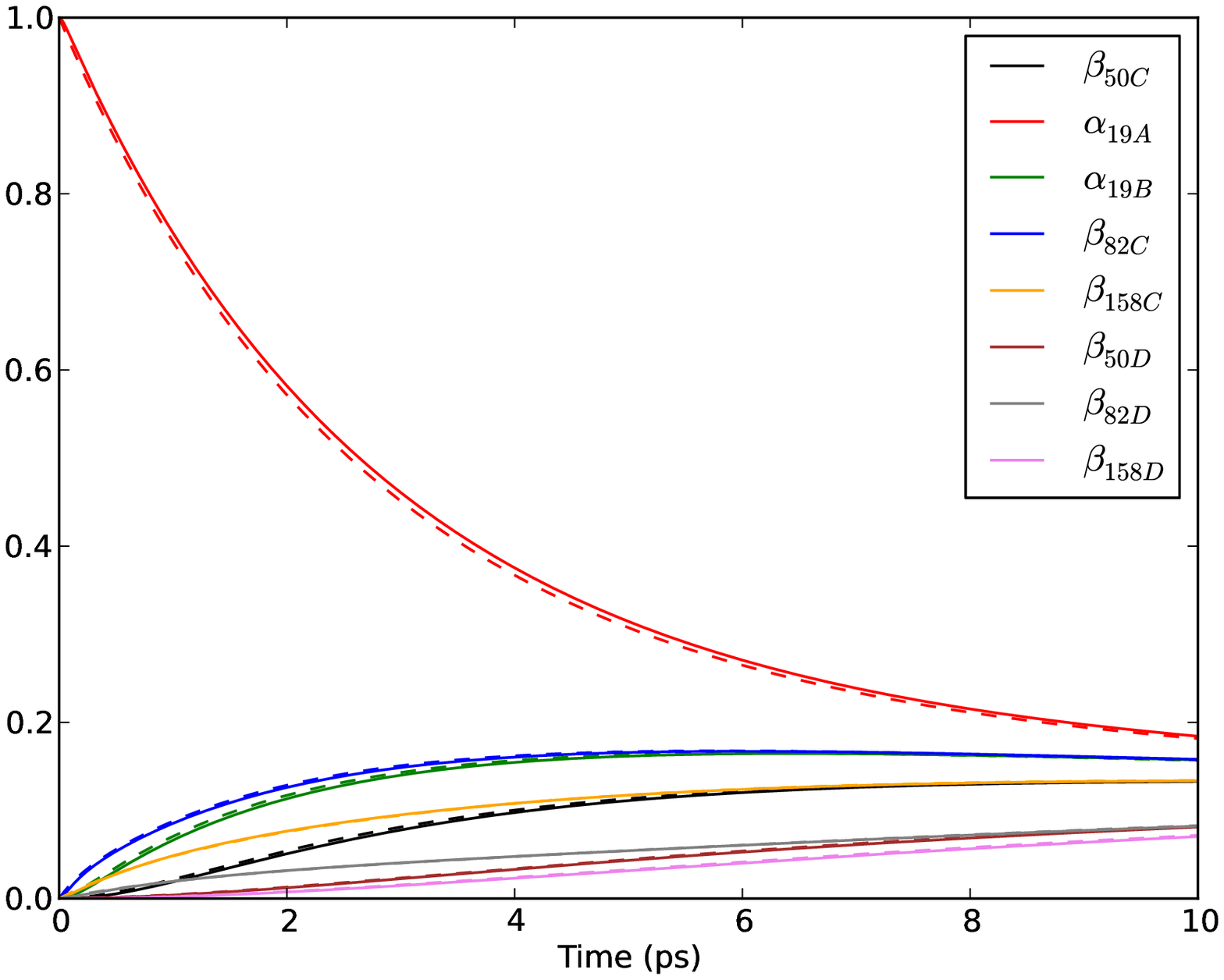}
\includegraphics[width=\columnwidth]{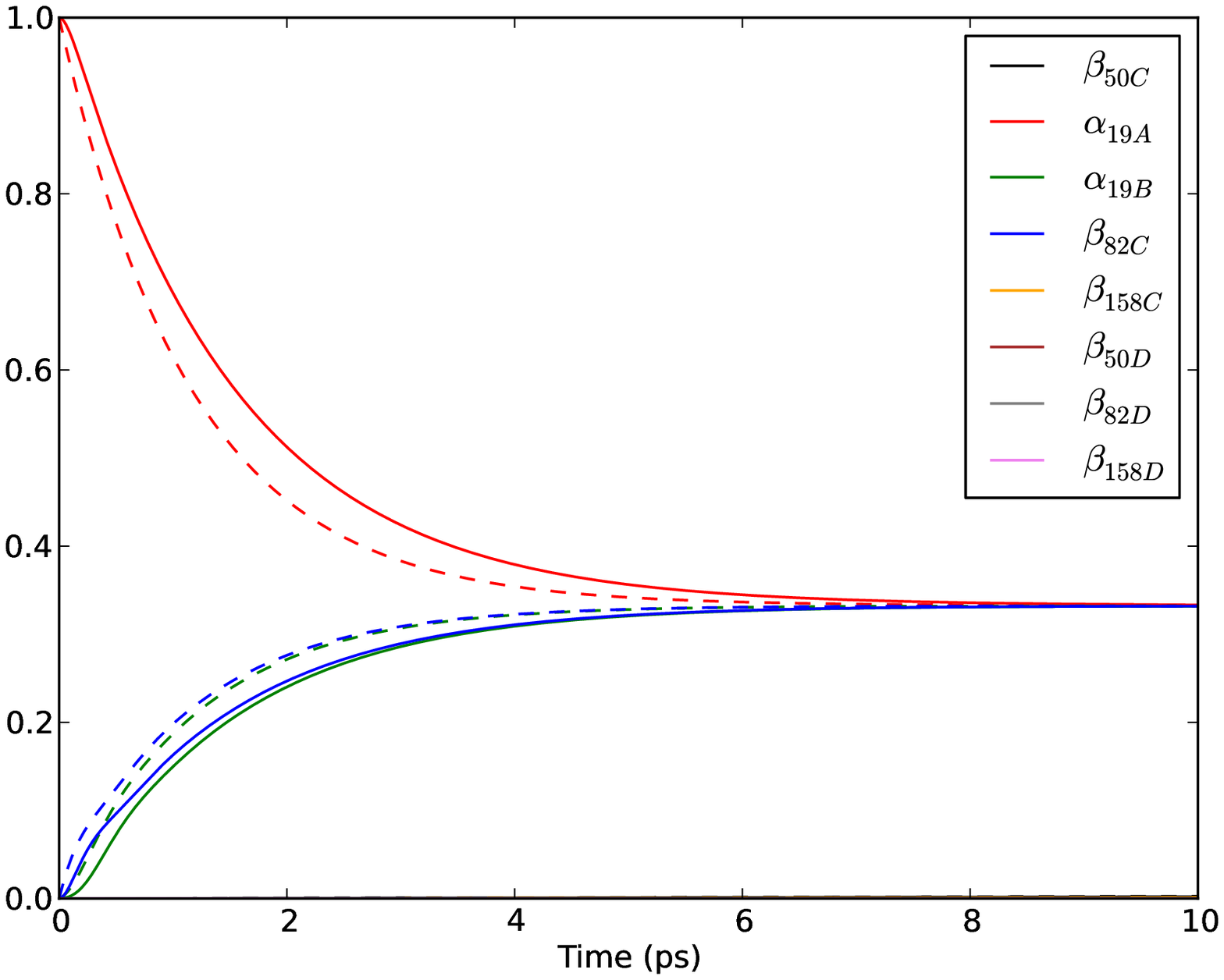}
\caption{Exciton populations vs time calculated at a) 300K b) 77K for initial
  state $\alpha_{19A}$.  Solid
  lines give populations calculated by Eq. \ref{eq:dpdt}, while dashed lines
  give populations calculated using the minimal exponential decay model.}
\label{fig:pops_vs_time_a19a}
\end{figure}

\begin{figure}
\includegraphics[width=\columnwidth]{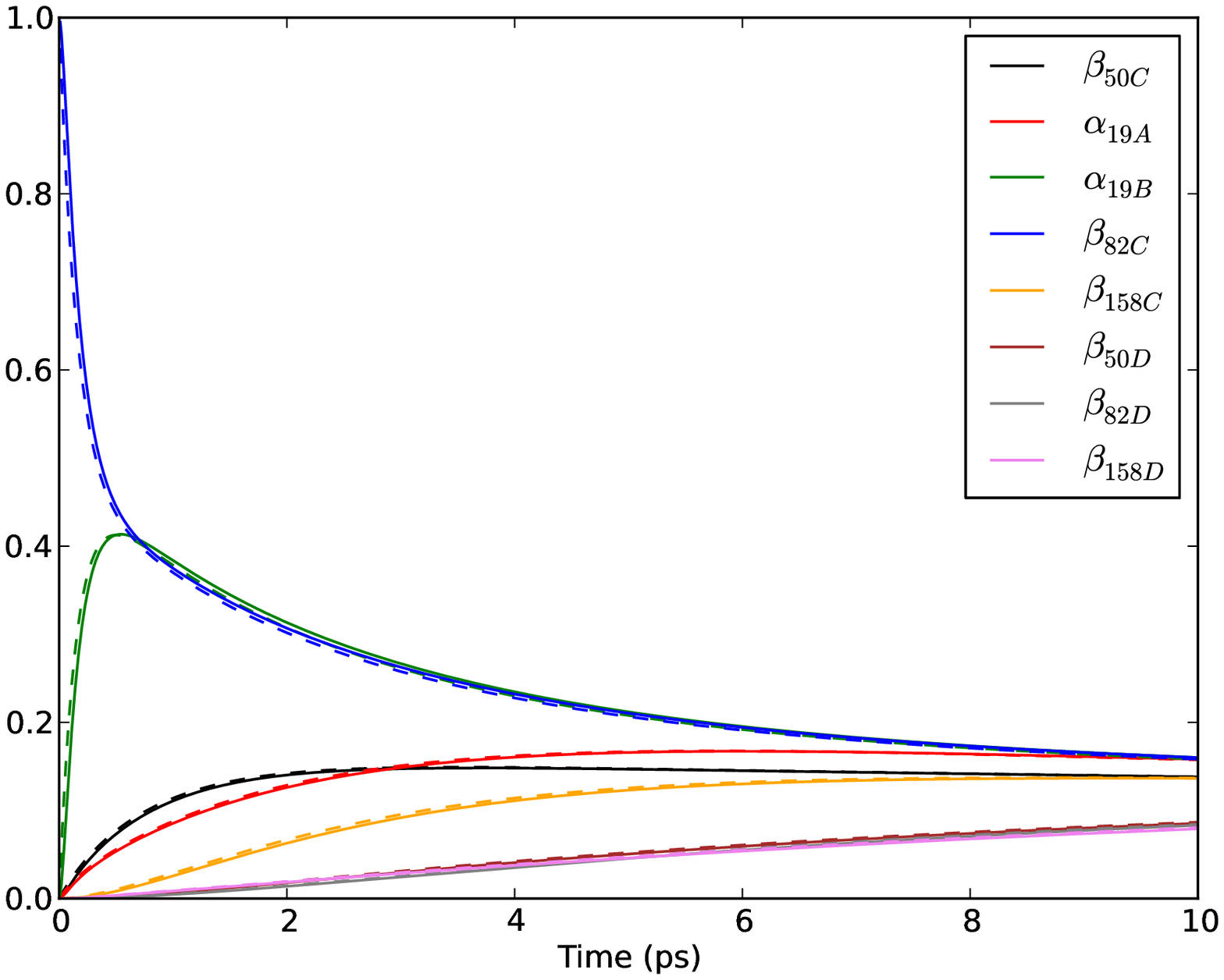}
\includegraphics[width=\columnwidth]{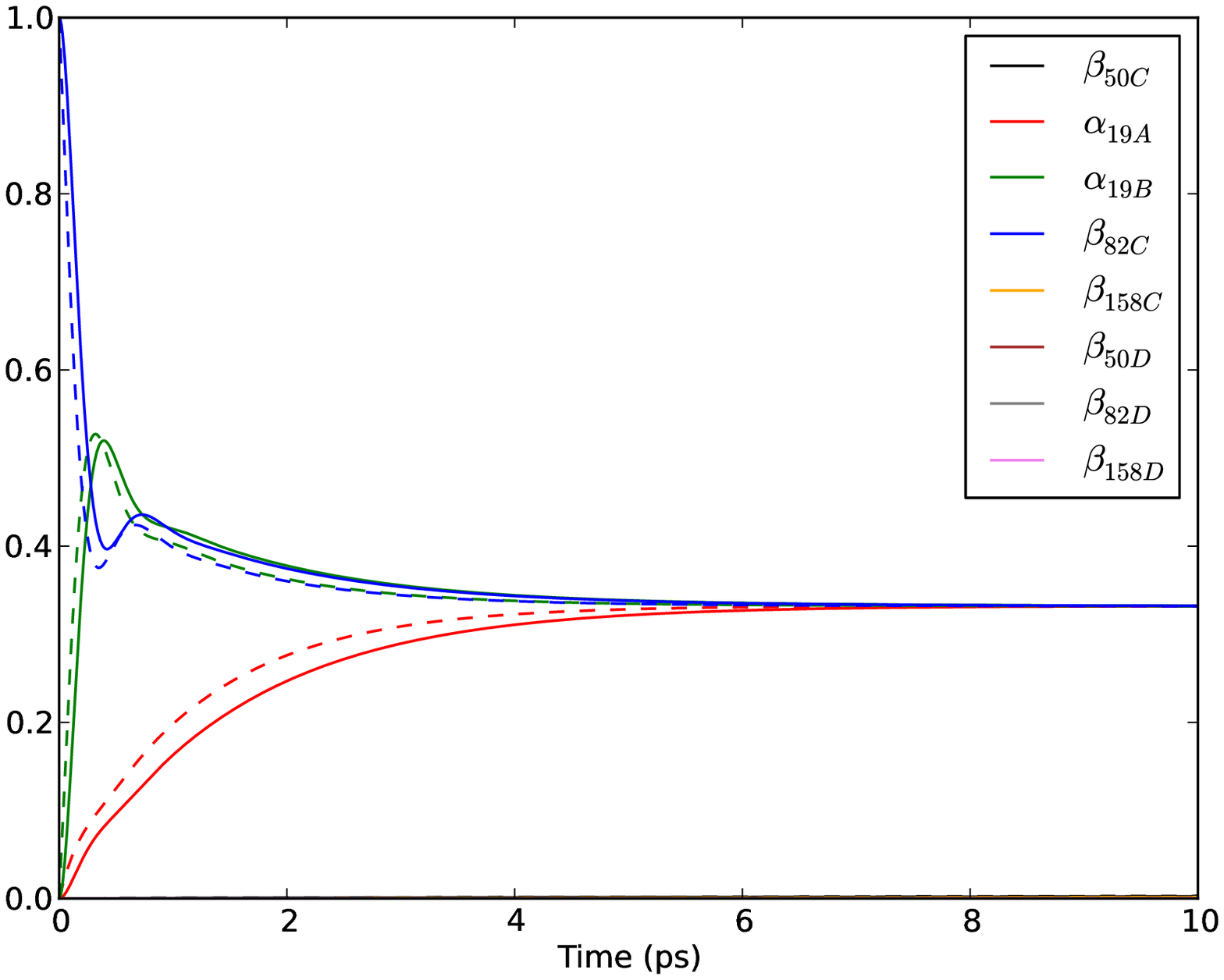}
\caption{Exciton populations vs time calculated at a) 300K b) 77K for initial
  state $\beta_{82C}$.  Solid
  lines give populations calculated by Eq. \ref{eq:dpdt}, while dashed lines
  give populations calculated using the minimal exponential decay model.}
\label{fig:pops_vs_time_b82c}
\end{figure}

\begin{figure}
\includegraphics[width=\columnwidth]{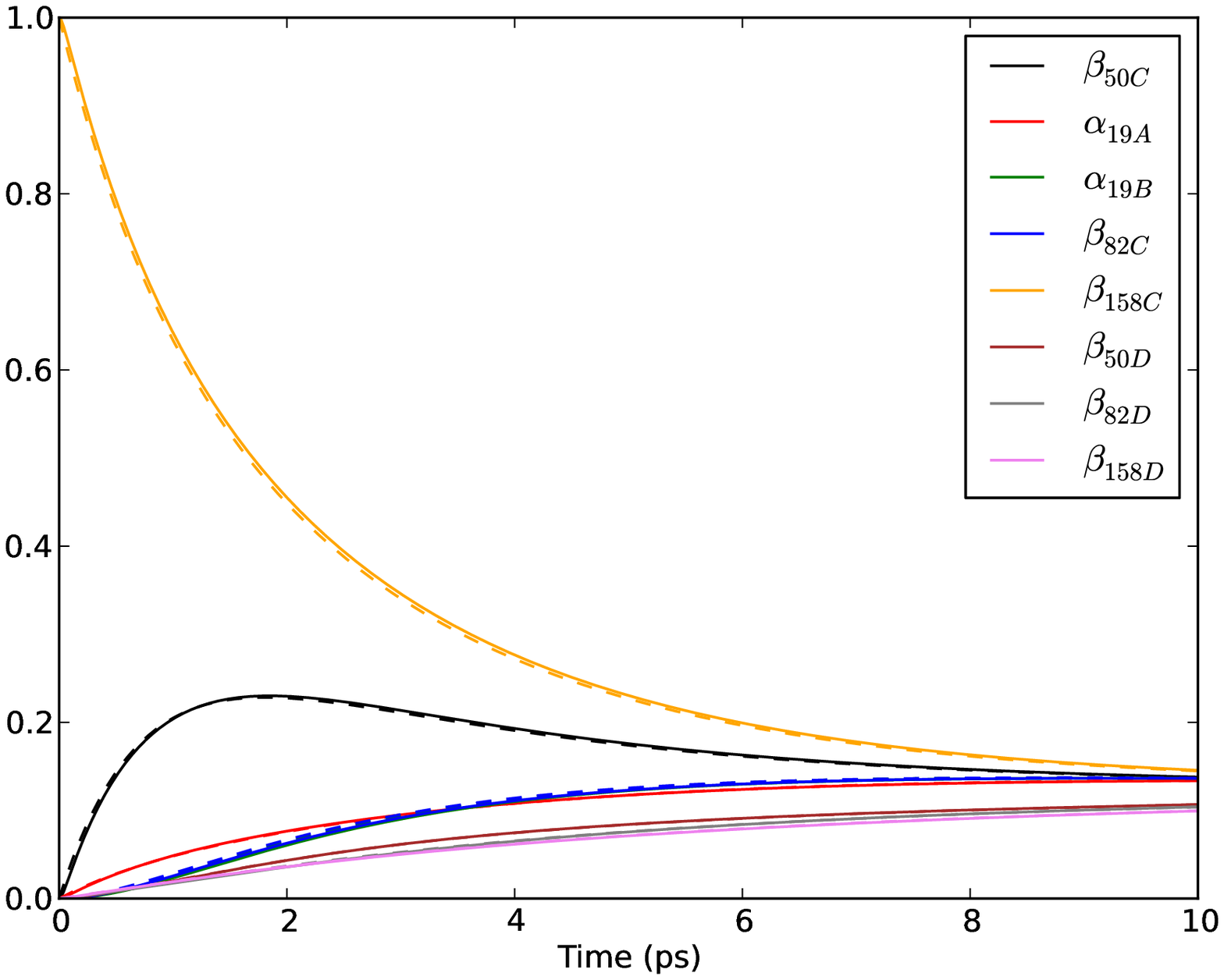}
\includegraphics[width=\columnwidth]{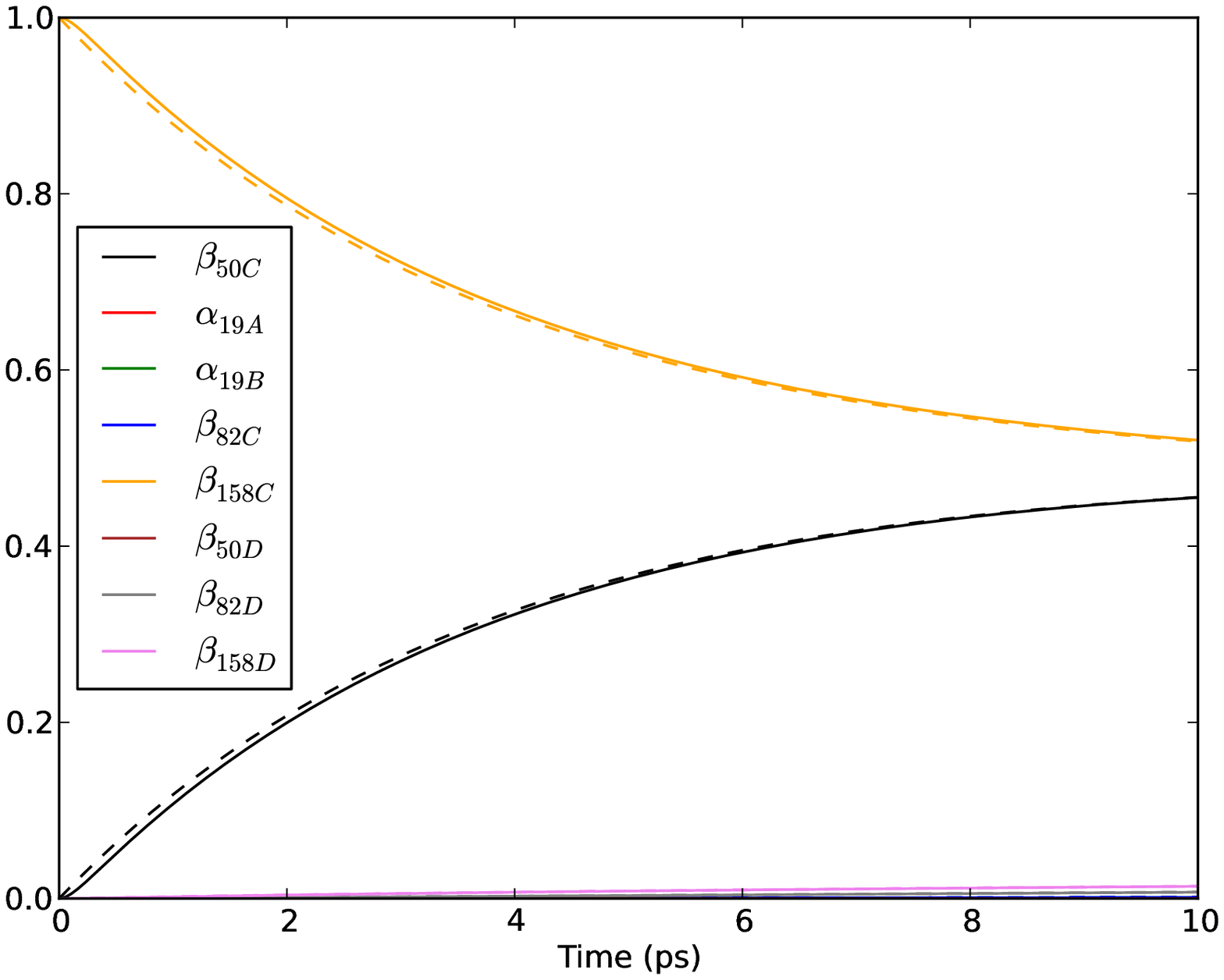}
\caption{Exciton populations vs time calculated at a) 300K b) 77K for initial
  state $\beta_{158C}$.  Solid
  lines give populations calculated by Eq. \ref{eq:dpdt}, while dashed lines
  give populations calculated using the minimal exponential decay model.}
\label{fig:pops_vs_time_b158c}
\end{figure}

\begin{figure}
\includegraphics[width=\columnwidth]{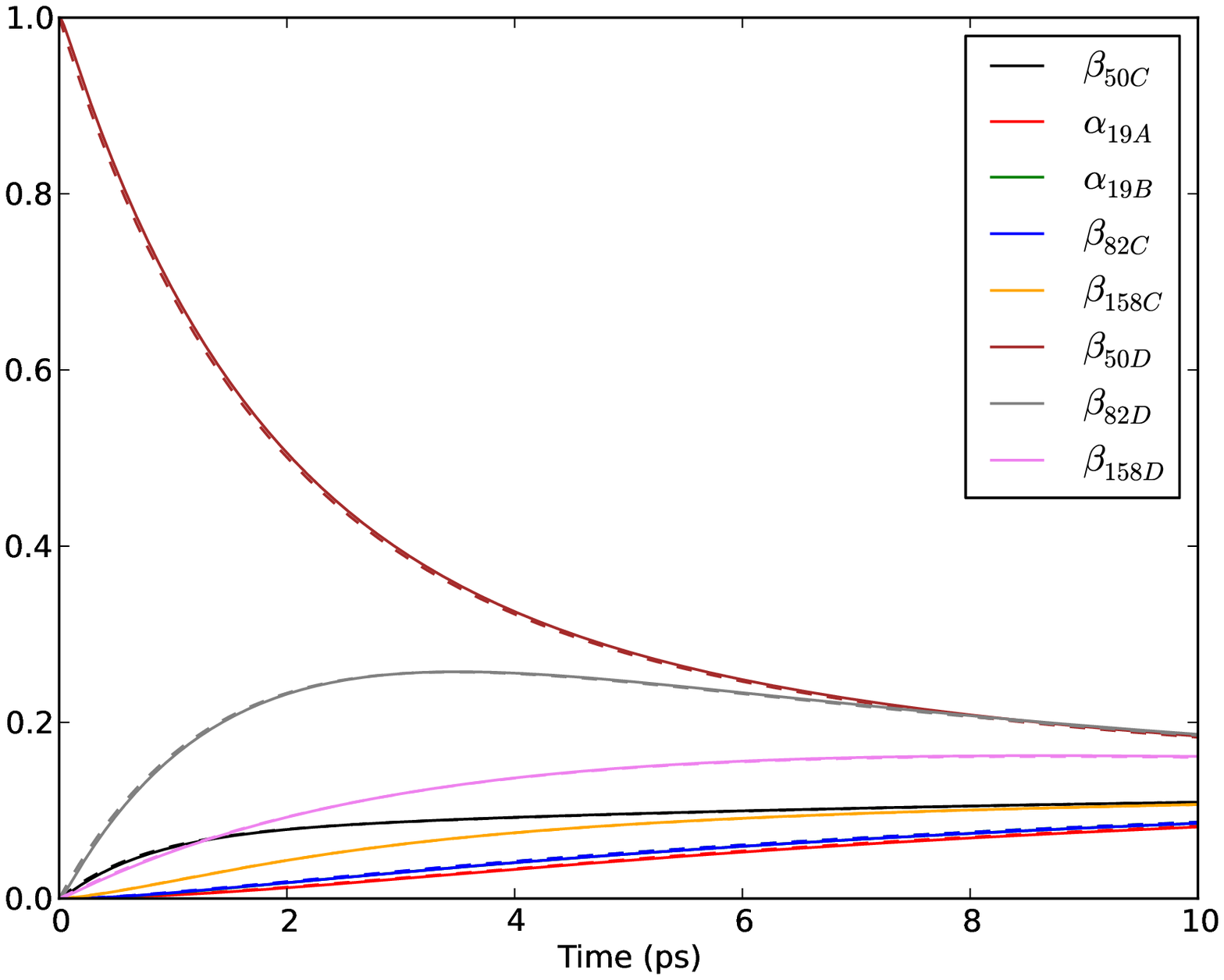}
\includegraphics[width=\columnwidth]{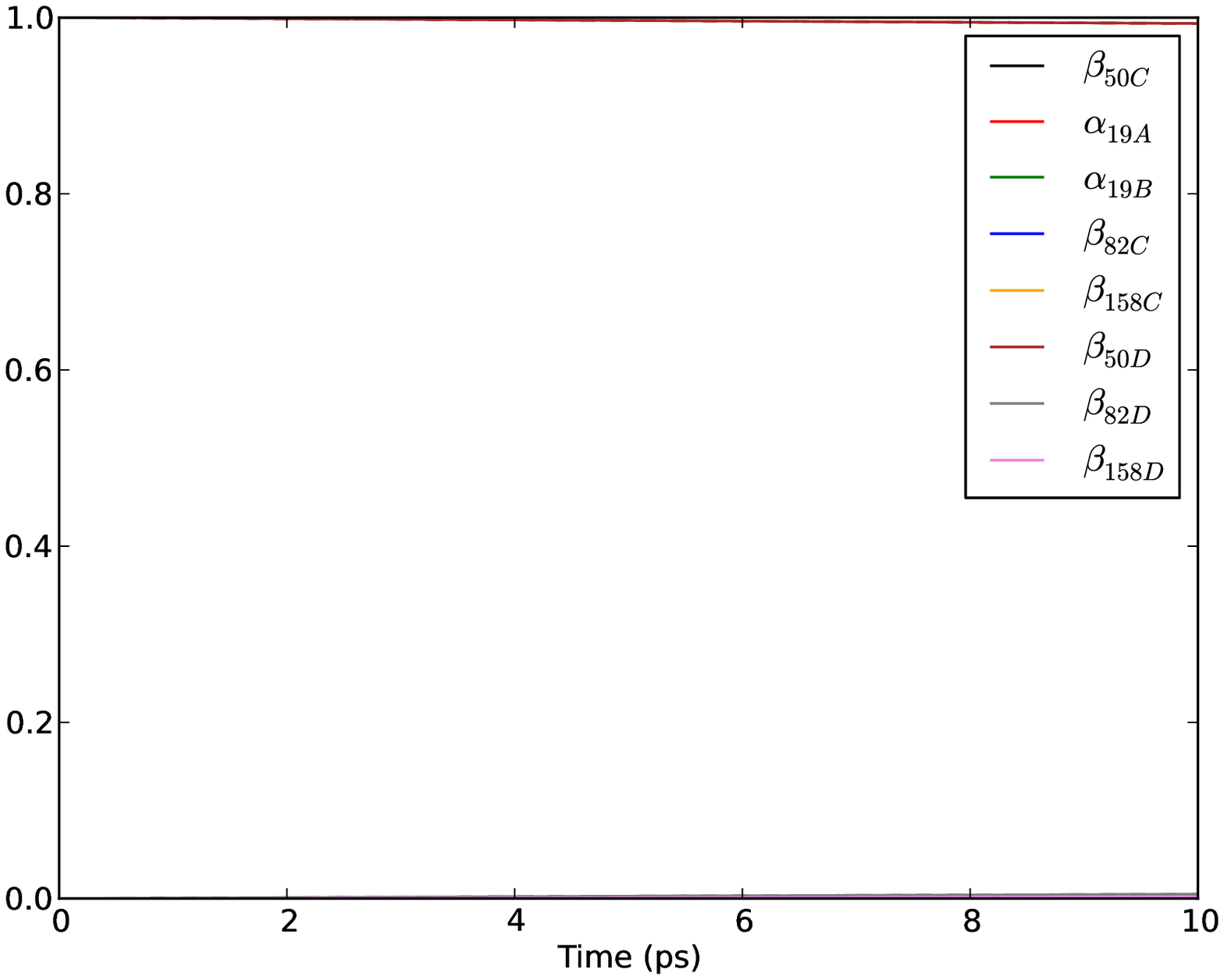}
\caption{Exciton populations vs time calculated at a) 300K b) 77K for initial
  state $\beta_{50D}$.  Solid
  lines give populations calculated by Eq. \ref{eq:dpdt}, while dashed lines
  give populations calculated using the minimal exponential decay model.}
\label{fig:pops_vs_time_b50d}
\end{figure}

\begin{figure}
\includegraphics[width=\columnwidth]{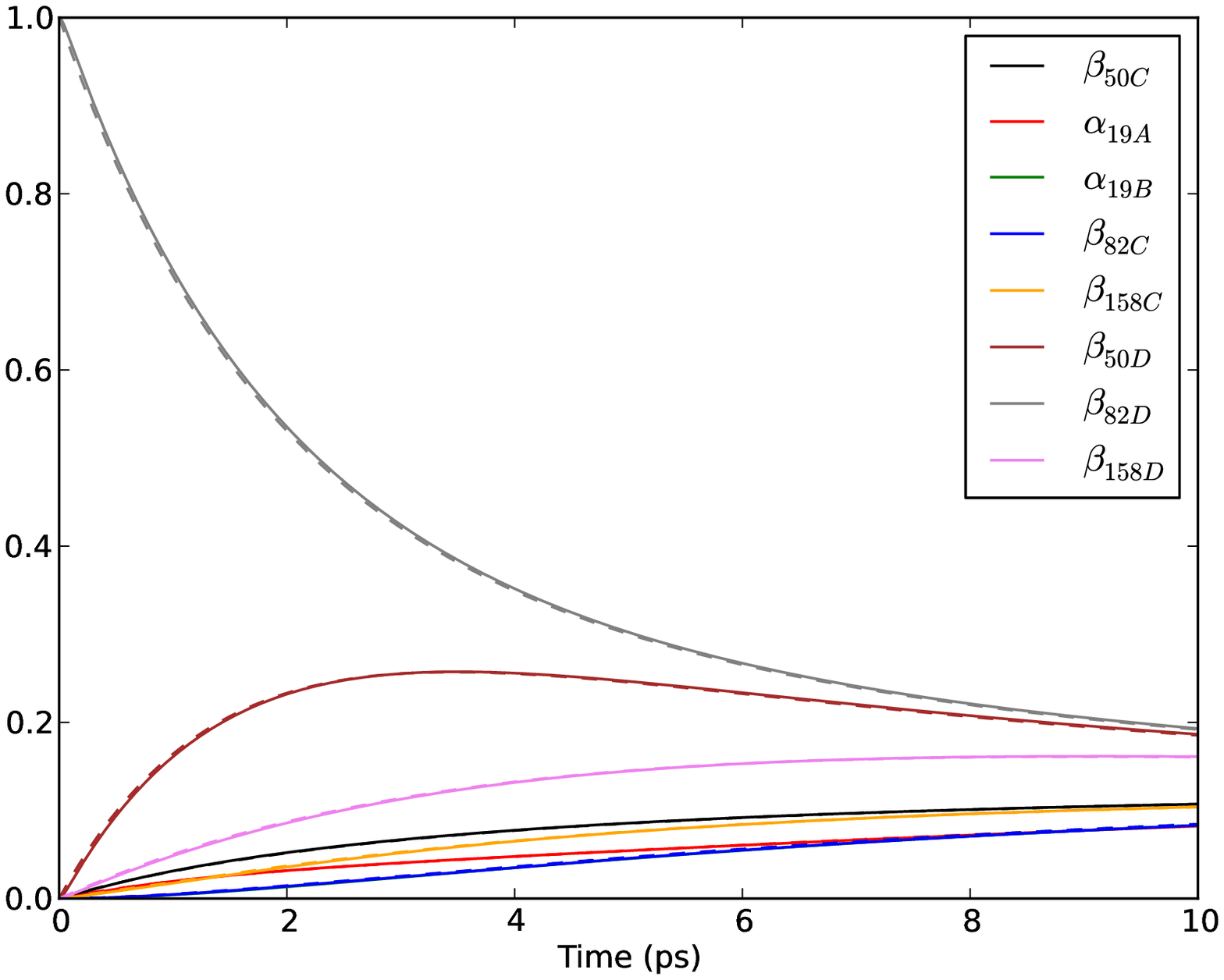}
\includegraphics[width=\columnwidth]{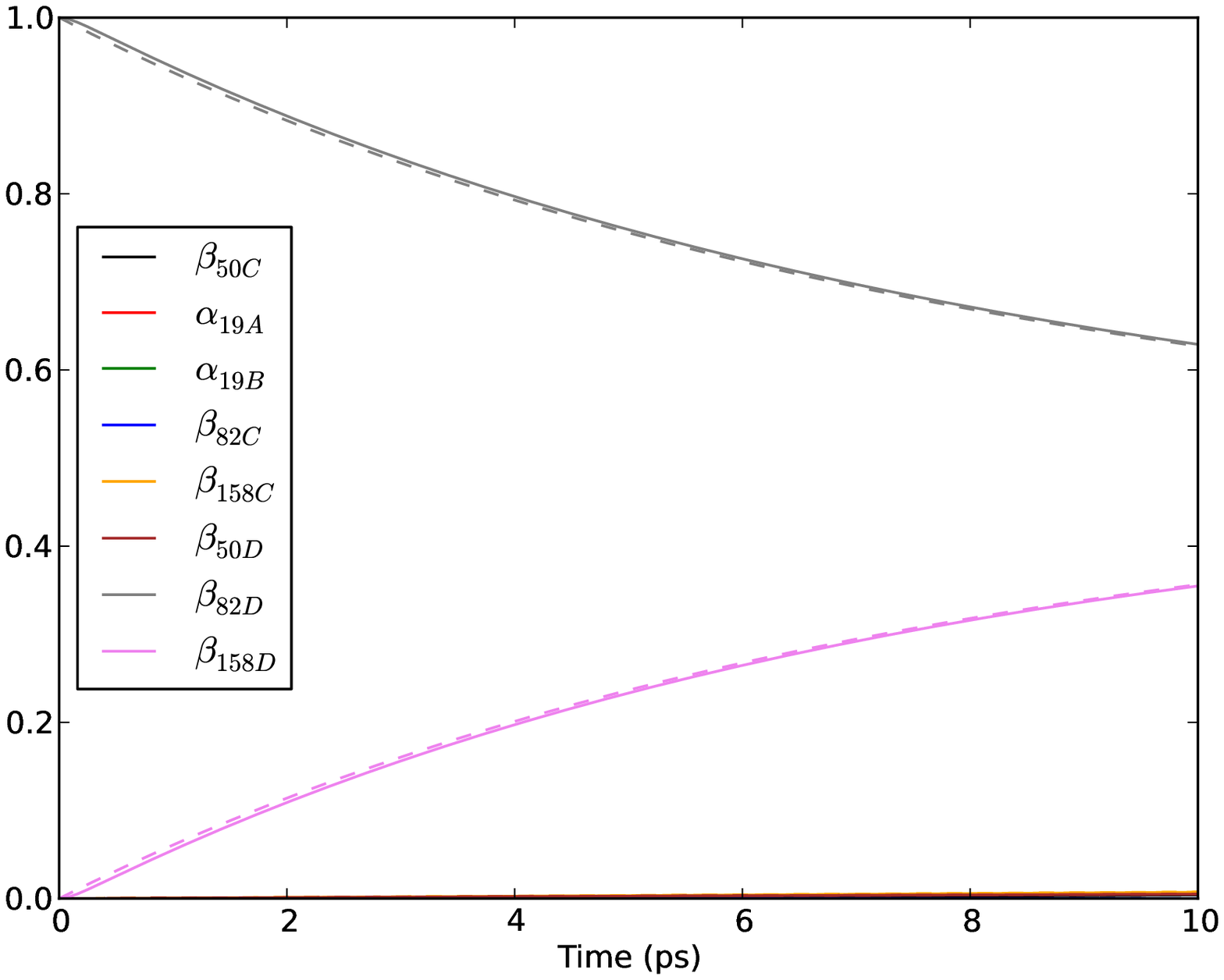}
\caption{Exciton populations vs time calculated at a) 300K b) 77K for initial
  state $\beta_{82D}$.  Solid
  lines give populations calculated by Eq. \ref{eq:dpdt}, while dashed lines
  give populations calculated using the minimal exponential decay model.}
\label{fig:pops_vs_time_b82d}
\end{figure}

\begin{figure}
\includegraphics[width=\columnwidth]{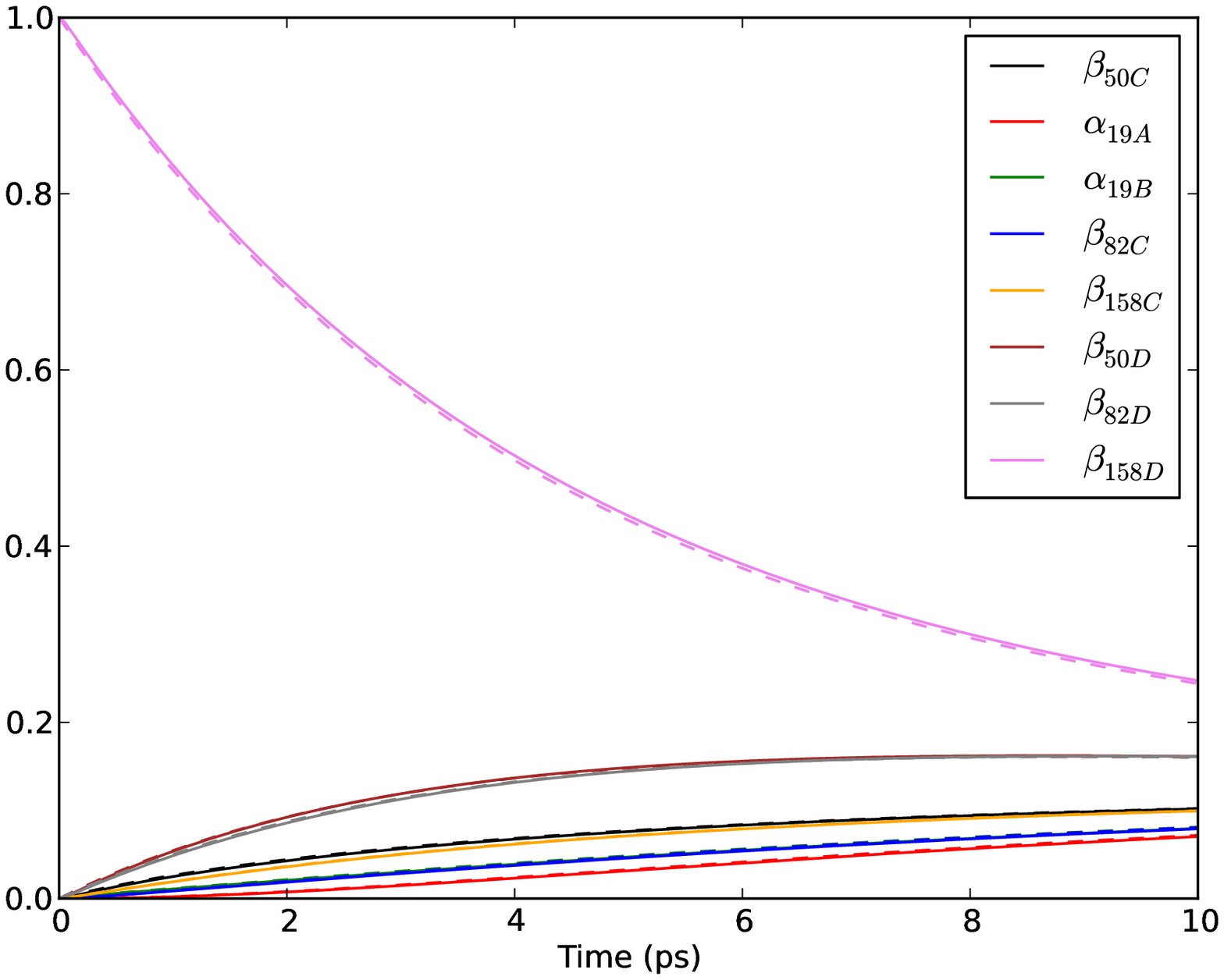}
\includegraphics[width=\columnwidth]{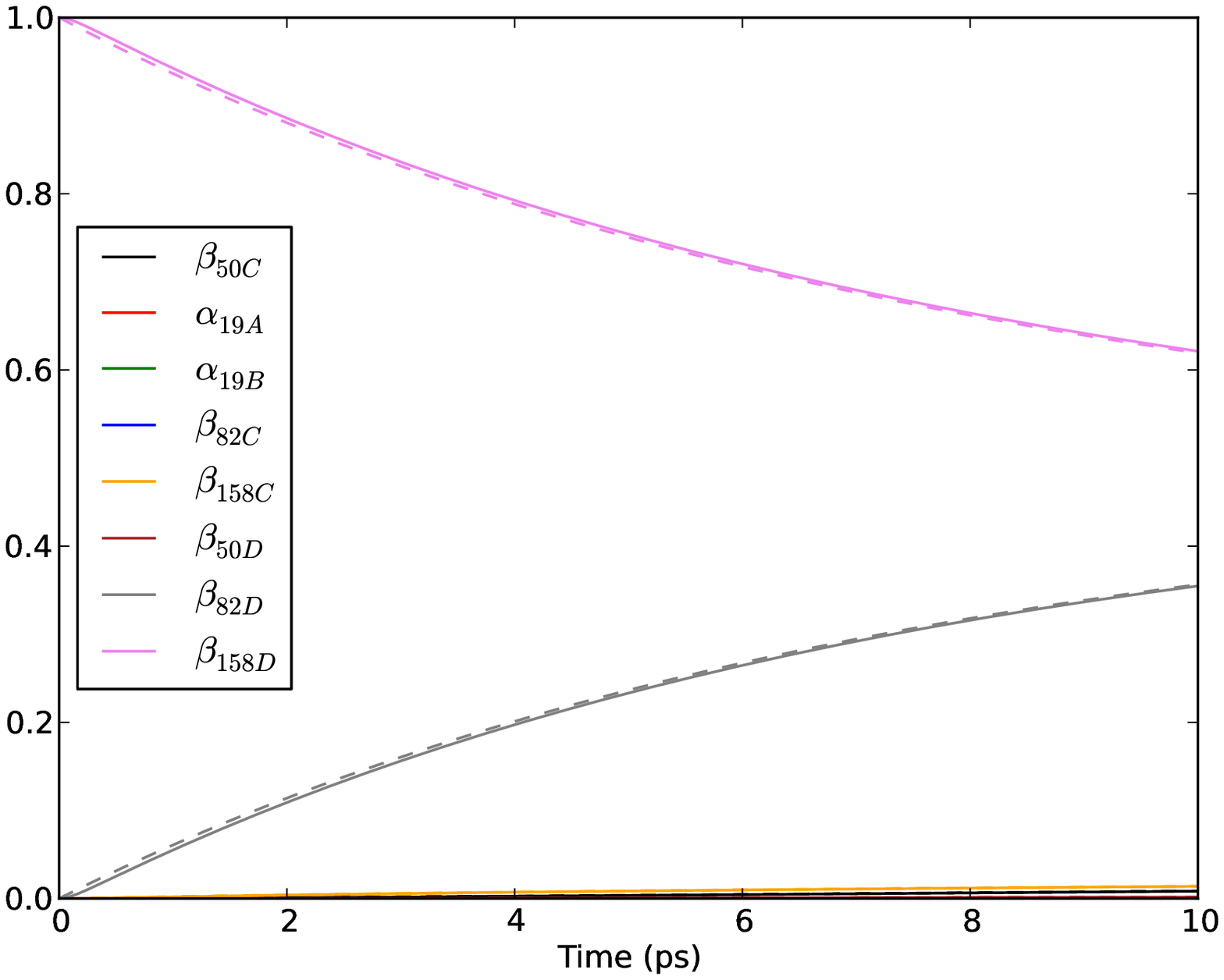}
\caption{Exciton populations vs time calculated at a) 300K b) 77K for initial
  state $\beta_{158D}$.  Solid
  lines give populations calculated by Eq. \ref{eq:dpdt}, while dashed lines
  give populations calculated using the minimal exponential decay model.}
\label{fig:pops_vs_time_b158d}
\end{figure}

The close agreement between the minimal model and the full density matrix
propagation is shown in Figures
%\ref{fig:pops_vs_time_a19b,fig:pops_vs_time_b50c,fig:pops_vs_time_a19a,
%  fig:pops_vs_time_b82c,fig:pops_vs_time_b158c,fig:pops_vs_time_b50d,
%  fig:pops_vs_time_b82d,fig:pops_vs_time_b158d}
\ref{fig:pops_vs_time_a19b}, \ref{fig:pops_vs_time_b50c},
\ref{fig:pops_vs_time_a19a}, \ref{fig:pops_vs_time_b82c},
\ref{fig:pops_vs_time_b158c}, \ref{fig:pops_vs_time_b50d},
\ref{fig:pops_vs_time_b82d}, and \ref{fig:pops_vs_time_b158d}, which compare
the two models for initial conditions with the excitons localized on each
individual chromophore.  The largest departure between the two models is
observed for $\alpha_{19A}$, $\beta_{82C}$ and $\alpha_{19B}$ initial
conditions at 77K, which correspond to the closed curve of efficient transfer
pathways previously identified.

%%%%%%%%%%%%%%%%Extra figures

The close agreement between the two models explains why the experimental
transfer rates show good agreement with the pairwise transfer rates, and
indicates that the information contained in the long lived antisymmetric
coherence terms, which are not included in the exponential decay model, does
not play a vital role in the transfer of exciton population.  Although the
antisymmetric coherence terms show the same picosecond timescales for decay as
the population imbalances, the lack of efficient, closed cycles in the PE545
complex leave little opportunity for interference.  In addition, all initial
conditions settle rapidly to the same equilibrium distribution at 300K, so
that any "quantum speedup" would offer only transient advantages.  For these
reasons, it appears that the importance of coherence terms in PE545 is dynamic
rather than informational -- coherence terms are important because they are
required to give the correct transfer rates between chromophores, but the
information contained in these terms does not appear to play an essential role
in the transfer of the exciton population to the radiating DBV chromophore.

\section{Conclusions}
This paper has presented a new theory of coherent dynamics in photosynthetic
complexes.  Strong interaction with a reservoir causes rapid thermalization of
vibrational states, allowing equations of motion for slowly varying electronic
coefficients to be derived.  By incorporating thermodynamic quantities into
the definition of a thermalized density matrix and an effective coupling,
these equations of motion are reduced to the Haken, Reineker, Strobl form.
Unlike the original HRS model, the incorporation of thermodynamic information
causes the system to settle to the correct thermodynamic equilibrium rather
than to equal population of all states.  The resulting theory gives good
theoretical lineshapes for absorption, fluorescence and circular dichroism for
PE545 at 300K and 77K, and yields exciton transfer times which agree with
those found by EADS experiments.  For both kinds of experiments, agreement is
better at 300K than at 77K.

Long lived coherences between chromophores, such as those observed in
\cite{lee2007coherence,collini2010coherently,gregory2007evidence,
  savikhin1997oscillating}, are simply explained in terms of an overdamped
harmonic oscillator, in which the strength of the damping is proportional to
the temperature and the effective coupling decreases exponentially with the
excitation energy separation between two chromophores.  Because of this,
coherence terms may survive for arbitrarily long times, even in the high
temperature limit for a system which interacts strongly with its environment.

Somewhat surprisingly, the long lifetime for survival of coherence terms
appears to have a minimal effect on the transfer of exciton population in the
PE545 antenna complex.  This can be understood as resulting from the
relatively sparse network of efficient transfer pathways in this complex,
which has the effect of limiting interference between competing pathways.  A
minimal exponential decay model describes the flow of exciton population
through this network accurately at 300K, when this network has no closed
cycles, but departs from the full density matrix calculation at 77K, when such
a cycle allows for interference between multiple pathways through the
complex.  It is possible that other photosynthetic molecules, with functions
more sophisticated than the simple direction of exciton population to the
lowest energy chromophore, may make more sophisticated use of this
information.

The method of thermalized reduction of the density matrix introduced in this
paper allows the derivation of quantum coherent equations of motion in a
regime of high temperature and strong interaction with the surrounding
environment which is often considered inimical to quantum mechanical
behavior.  Rather than being destroyed rapidly by interaction with the
reservoir, coherence terms may persist or even be preserved by interaction
with the reservoir.  The conditions necessary for reduction -- that the
reduced degrees of freedom thermalize rapidly with respect to the evolution of
the unreduced degrees of freedom -- are satisfied in the limit of strong
interaction with the reservoir, while the thermodynamic weighting of the
effective coupling matrix may yield slow evolution of the unreduced degrees of
freedom even when the bare coupling is relatively strong.  The simplicity of
the conditions necessary for the thermalized reduction procedure show that
quantum mechanical behavior does not require low temperatures or elaborate
isolation from the surroundings to be observed.

\section{Acknowledements}
This work was supported in part by the SAW grant of the Liebniz society.

\end{document}